\documentclass[a4paper,fleqn,usenatbib]{mnras}

\usepackage{newtxtext,newtxmath}

\usepackage[T1]{fontenc}
\usepackage{ae,aecompl}

\usepackage{graphicx}	
\usepackage{amsmath}	
\usepackage{amssymb}	
\usepackage{wasysym}

\title[Photoevaporation of a Trappist-1 disc]{Where can a Trappist-1 planetary system be produced? }
\author[T. J. Haworth et al.]
{\parbox{\textwidth}{Thomas J. Haworth$^{1}$\thanks{E-mail: \texttt{t.haworth@imperial.ac.uk}}, Stefano Facchini$^{2}$, Cathie J. Clarke$^{3}$, Subhanjoy Mohanty$^{1}$
}\vspace{0.4cm}\\
\parbox{\textwidth}{$^{1}$ Astrophysics Group, Imperial College London, Blackett Laboratory, Prince
Consort Road, London SW7 2AZ, UK \\
$^{2}$ Max-Planck-Institut f\"ur Extraterrestrische Physik, Giessenbachstrasse 1, 85748 Garching, Germany \\
$^{3}$ Institute of Astronomy, Madingley Rd, Cambridge, CB3 0HA, UK \\
}}

\begin{document}

\date{Accepted ???. Received ???; in original form ???}

\pagerange{\pageref{firstpage}--\pageref{lastpage}} \pubyear{2016}

\maketitle
\label{firstpage}

\begin{abstract}
We study the evolution of protoplanetary discs that would have been precursors of a Trappist-1 like system under the action of accretion and external photoevaporation in different radiation environments. Dust grains swiftly grow above the critical size below which they are entrained in the photoevaporative wind, so although gas is continually depleted, dust is resilient to photoevaporation after only a short time.  This means that the ratio of the mass in solids (dust plus planetary) to the mass in gas rises steadily over time. Dust is still stripped early on, and the initial disc mass required to  produce the observed $4\,M_{\oplus}$ of Trappist-1 planets is high. For example, assuming a \cite{2008ApJ...675.1361F} distribution of UV fields, typical initial disc masses have to be $>30$\,per cent the stellar (which are still Toomre $Q$ stable) for the majority of similar mass M dwarfs to be viable hosts of the Trappist-1 planets. Even in the case of the lowest UV environments observed, there is a strong loss of dust due to photoevaporation at early times from the weakly bound outer regions of the disc. This minimum level of dust loss is a factor two higher than that which would be lost by accretion onto the star during 10 Myr of evolution. Consequently even in these least irradiated environments, discs that are viable Trappist-1 precursors need to be initially massive ($>10$\,per cent of the stellar mass).
\end{abstract}


\section{introduction}
The recent discovery of 7 planets ranging from a quarter Earth to around 1.5 Earth masses, packed within 0.06\,AU of the 0.08\,M$_\odot$ M dwarf Trappist-1 \citep{2017Natur.542..456G, 2017arXiv170404290W} has unsurprisingly drawn considerable attention. In particular the habitability of planets in this system is uncertain, even though they reside in the ``habitable zone'' \citep[e.g.][]{2017MNRAS.469L..26O, 2017ApJ...844...19A, 2017arXiv170505535D, 2017MNRAS.464.3728B}. It has also raised some interesting questions regarding the formation mechanism for such a system. For example, assuming an initial dust-to-gas mass ratio of $10^{-2}$, total disc mass (approximately the gas mass) $M_g$, dust mass $M_d$, stellar mass $M_*=0.08$\,M$_\odot$ and planetary masses $\sum m_p=4\,M_{\oplus}$ \citep{2017arXiv170404290W}  the minimum planet formation efficiency (conversion of solids into planets) is 
\begin{equation}
	\eta \equiv  10^2\left(\frac{\sum m_p}{M_{d}}\right)\% = 10^4\left(\frac{\sum m_p}{M_g}\right)\% \approx \left(\frac{M_*}{M_g}\right)\%. 
	\label{pfe}
\end{equation}
where the last approximate equality uses the observed total planetary mass and stellar mass
in Trappist-1. This is necessarily high given that canonically $M_g < 0.1 M_*$. Furthermore, some mechanism is required to concentrate the solids for planet formation and also to produce the planets in the observed compact orbital configuration.

\cite{2017A&A...604A...1O} recently proposed a mechanism whereby the Trappist-1 planets are formed both quickly and efficiently. Grains grow to pebble size (which is associated with rapid
radial drift) at a time that increases radially outward through the disc. These pebbles then drift inwards to the water snow line. Streaming instabilities \citep[e.g.][]{2005ApJ...620..459Y, 2007Natur.448.1022J, 2016arXiv161107014Y} then produce cores at that location. Once the cores form they migrate inwards, still accreting as they do so, and settle into mean motion resonance (MMR). The lack of MMR for the innermost two planets of Trappist-1 is explained by internal clearing of the disc at the orbits of those planets \citep[although][have noted that this doesn't fully explain the 3-body Laplace resonance between the inner three planets]{2017arXiv171107932P}. Nevertheless, a mechanism of this form has the high efficiency expected to be required for the formation of a Trappist-1 system and also produces planets relatively quickly, making it (or some similar mechanism) a promising formation scenario.

It is unlikely that all the raw solid material
in the disc is available for conversion into planets because
of the action of a variety of dispersal mechanisms, including accretion, internal winds driven by photoevaporation and/or magnetic fields, and external photoevaporation by nearby stars. Because the streaming instability is sensitive to the solids-to-gas mass ratio  it could possibly be promoted by \textit{internal} photoevaporation of the disc by the host star \citep[see][]{2017ApJ...839...16C, 2017MNRAS.472.4117E}. In the present case of very low mass stars, internal photoevaporation is however likely to be subordinate
to the effect of the environment and we do not consider this further here.

The impact of external photevaporation on the evolution of protoplanetary discs is becoming increasingly apparent, across a wider range of UV field strengths than previously expected. For example proplyds (the cometary products of externally photoevaporating circumstellar discs) have long been observed in $\sim10^5$\,G$_0$\footnote{G$_0$ is the Habing unit, which is $1.6\times 10^{-3}$ erg cm$^{-2}$ s$^{-1}$ over the wavelength range ($912$\AA$<\lambda<2400$\AA)  \cite{1968BAN....19..421H}} environments in the vicinity of O stars \citep{1998AJ....115..263O, 2003ApJ...587L.105S,  2012ApJ...746L..21W}, but have also recently been detected in environments where the UV field strength is as weak as $\sim3000$\,G$_0$ \citep{2016ApJ...826L..15K}. Given that the median UV field environment is of order $1000$\,G$_0$ \citep{2008ApJ...675.1361F} we hence expect that a large fraction of stellar systems could have been influenced by photoevaporation.

There is also observational evidence that both the incidence
\citep{2017A&A...602A..10G} and the mass  \citep{2014ApJ...784...82M, 2017AJ....153..240A} of discs
in clusters increases as a function of distance from the major
ultraviolet sources in the cluster (i.e. the O stars). Furthermore there is the theoretical expectation of effective photoevaporation in lower radiation regimes \citep{2004ApJ...611..360A, 2011PASP..123...14H, 2016MNRAS.457.3593F}. In particular, where the gravitational potential from the star is shallow, even very low UV fields can drive significant mass loss, as is expected for the large disc in IM Lup which is being irradiated by a field of only $\sim4$\,G$_0$ \citep{2016ApJ...832..110C, 2017MNRAS.468L.108H}. This therefore raises the possible expectation that a disc around a very low mass star such as Trappist-1 may be subject to substantial photoevaporation if disc material is only quite weakly bound \citep[there are some observed instances of evaporating discs around low mass stars, e.g.][]{2016ApJ...833L..16F}.

In this paper we study the external photoevaporation of star-disc precursors to Trappist-1 like systems. We aim to determine the mass loss rates of gas and dust from such a disc and hence gauge the impact on the evolution and planet formation potential of a Trappist-1 analogue star-disc system.


\section{Numerical method}
\label{sec:method}

\subsection{Overview}

We begin with an overview of our general method, after which we will provide more details of the contributing components. 

Our models are 1D viscous evolutionary calculations that account for accretion onto the central star, as well as photoevaporative mass loss from the disc outer edge. Although the accretion and viscous evolution is not numerically intensive, and such calculations have been done many times before \citep[e.g.][]{2007MNRAS.376.1350C, 2013ApJ...774....9A, 2015ApJ...815..112K, 2017MNRAS.468.1631R}, photoevaporative mass loss rates from externally irradiated discs are expensive to compute and therefore cannot be done on-the-fly as a disc viscously evolves. We therefore take the approach of pre-computing the mass loss rate of a disc about a Trappist-1 mass (0.08\,M$_{\odot}$) star in terms of the surface density $\Sigma_d$ at the disc outer edge $R_d$ for a disc irradiated by a UV field in units of $G_0$. We then interpolate from this grid of $\dot{M}_w(R_d, \Sigma_d, G_0)$ (where $w$ denotes ``wind'') during our viscous evolutionary calculations. In addition, we track the growth of grains in our disc, which affects the \textit{dust} mass loss rate $\dot{M}_{dw}$ since only grains below a certain size are entrained in the flow \citep{2016MNRAS.457.3593F}.

\begin{figure}
	\hspace{-0.5cm}
	\includegraphics[width=10cm]{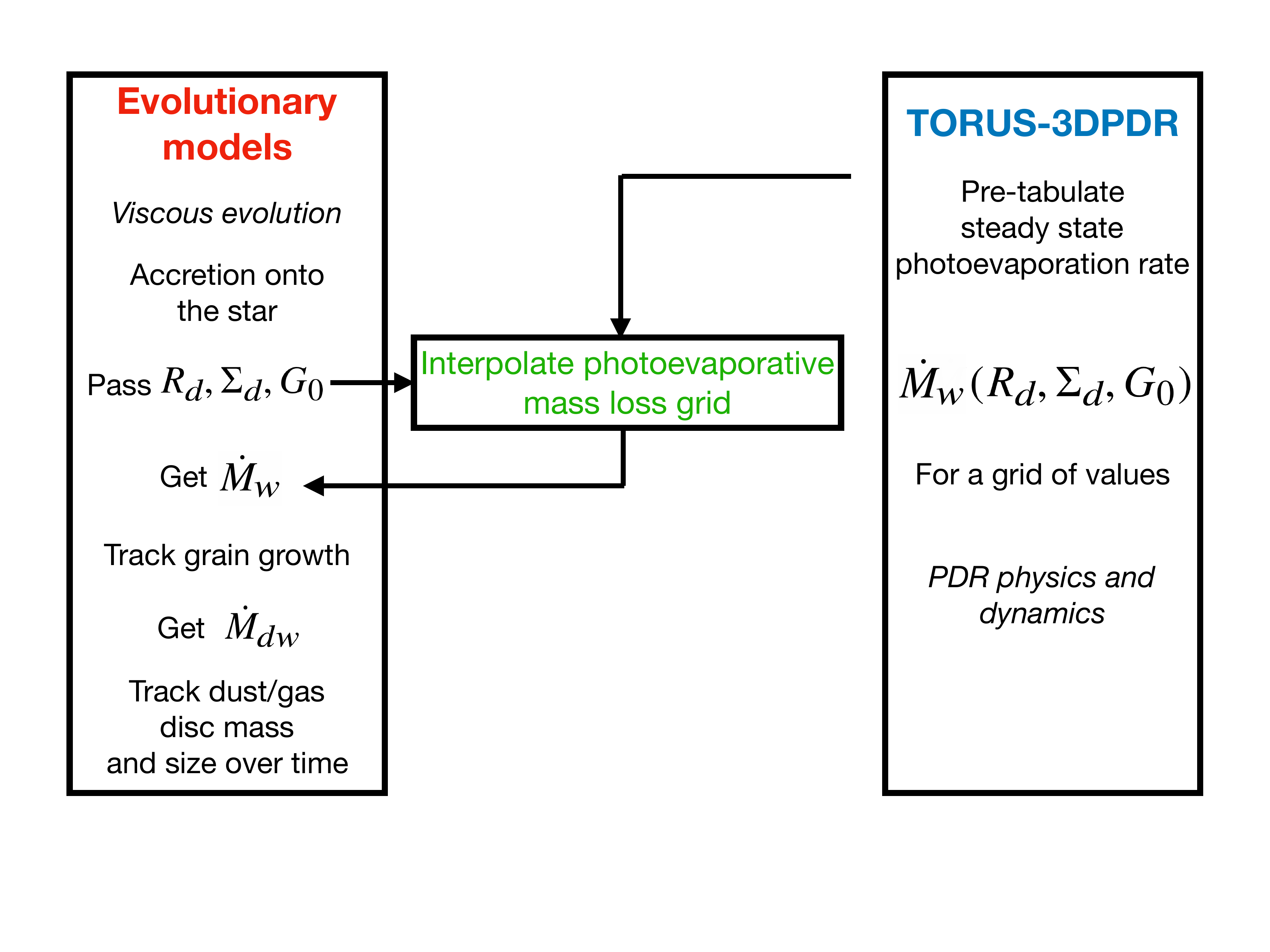}
	\vspace{-1cm}
	\caption{A schematic of our methodology. We pre-tabulate the computationally expensive photoevaporative mass loss rate as a function of the disc outer properties and UV field strength. This is then interpolated over to provide the mass loss rate in our viscous evolutionary models. }
	\label{methodSchematic}
\end{figure}

Using a grid of such viscous evolutionary models we study the disc mass/size evolution and its implications for planet formation. A pictorial overview of our scheme is shown in Figure \ref{methodSchematic}. {We summarise the key notation used in this paper in Table \ref{tab:symbols}. }

\subsection{Photoevaporative mass loss rates}
\label{mdotcalc}
We now discuss our technique for producing a grid of photoevaporative mass loss rates as a function of the properties at the disc outer edge. Note that as discussed above, these mass loss rates are computed prior to the viscous evolutionary models and are hence completely separate calculations. This separation is justified by
the fact that the timescale for secular evolution of the disc due
to combined viscous evolution and photoevaporation is large compared with
the timescale on which steady state photoevaporative flows are set up.

\subsubsection{Photochemical-dynamics with \textsc{TORUS-3DPDR}}
We use the \textsc{torus-3dpdr} photochemical-dynamical code to compute our grid of photoevaporative mass loss rates  \citep{2015MNRAS.453.2277H, 2015MNRAS.454.2828B, 2015MNRAS.448.3156H}. This code directly couples photodissociation region (PDR) microphysics with hydrodynamics, which has not been possible until recently owing to the complexity of the implementation and computational cost of the calculations \citep[e.g.][]{2016PASA...33...53H}. 

\textsc{torus-3dpdr} currently considers a reduced network that comprises 33 species and 330 reactions, tailored such that it provides temperatures to with $\sim10$\,per cent of the \textsc{3d-pdr} \citep{2012MNRAS.427.2100B} full chemical network (215 species, over 3000 reactions). It iteratively solves for the composition and temperature structure until convergence, using the approach detailed by \cite{2012MNRAS.427.2100B, 2015MNRAS.454.2828B}. The 33 species in the reduced network are presented in \cite{2016MNRAS.463.3616H}, but broadly includes atoms/ions/molecules comprised of H, He, C, O and Mg, as well as electrons, cosmic rays, PAHs and dust. 

The main coolants are lines from C, C+, O and CO, with some additional contribution from the dust. Heating processes include C ionisation, H$_2$ formation and photodissociation, FUV pumping, cosmic rays, turbulent and chemical heating and gas-grain collisions. The dominant heating contribution in a PDR can readily be photoelectric heating from atomic layers of polycyclic aromatic hydrocarbons \citep[PAHs, see e.g. Figure 2 of][]{2016MNRAS.457.3593F}, the abundance of which is highly uncertain, particularly towards the outer regions of discs \citep[e.g.][]{2006A&A...459..545G, 2010ApJ...714..778O, 2011ApJ...727....2P}. We therefore assume a negligible PAH abundance, which will mean that our heating of the disc, and hence mass loss rates, are likely lower limits. We assume a dust cross section in the photoevaporative wind of $\sigma_{\textrm{FUV}}=3\times10^{-23}$\,cm$^{-2}$, which is approximately the value found by \cite{2016MNRAS.457.3593F} when the maximum grain size in the disc is $\sim1$\,mm and the maximum grain size entrained in the flow is anywhere in the range $0.1\,\mu${m} -- $1$\,mm. 

\textsc{torus-3dpdr} uses a finite volume hydrodynamics scheme, and performs photochemical-dynamical simulations by iteratively computing PDR and hydrodynamical updates (i.e. via operator splitting). For the calculations in this paper the star dominates the gravitational potential, with self-gravity being negligible. We hence just consider a point source potential from a 0.08\,M$_{\odot}$ star. In  \cite{2016MNRAS.463.3616H} it was tested in the context of external photoevaporation models through comparison with the semi-analytic models of \cite{2016MNRAS.457.3593F}.

\medskip

\subsubsection{External disc photoevaporation}
To produce a grid of $\dot{M}_w(R_d, \Sigma_d, G_0)$ we require a series of steady state external photoevaporation models for different surface densities ($\Sigma_d$), radii ($R_d$) and UV fields ($G_0$). These have to encapsulate the range of conditions that will arise in our grid of viscous evolutionary models. To do so we set up a series of disc surface density profiles which we truncate, as described below. 

Our mass loss rate calculations are very similar to those of \cite{2016MNRAS.463.3616H, 2017MNRAS.468L.108H} only tailored to a Trappist-1 like system.  They are 1D, assuming that mass loss is predominantly driven from the disc outer edge \citep[e.g.][]{2004ApJ...611..360A, 2016MNRAS.457.3593F}. The argument for this simplification is two-fold. Firstly material at the disc outer edge is less gravitationally bound to the star, making it easier to liberate in a wind. Secondly the volume density in a disc drops off more rapidly vertically than radially outward, so there is a larger mass reservoir to feed the photoevaporative wind at the disc outer edge. The mass loss rate from these 1D models assumes that the flow is spherical from the disc outer edge, that is
 \begin{equation}
	\dot{M}_w = 4\pi R^2 \rho \dot{R} \mathcal{F}
\end{equation}
where  $\dot{R}$  is the flow velocity, with density $\rho$ at distance $R$, and $\mathcal{F}$ is the fraction of solid angle subtended by the disc outer edge $R_d$, which \cite{2004ApJ...611..360A} define as
\begin{equation}
	\mathcal{F} = \frac{H_d}{\sqrt{H_d^2+R_{d}^2}}
\end{equation} 
where $H_d$ is the scale height at this disc outer edge. 

In the calculations themselves we set up a \cite{1974MNRAS.168..603L} profile with $\gamma=1$, which yields
\begin{equation}
	\Sigma(R) = \Sigma_c\left(\frac{R}{R_c}\right)^{-1}\exp\left(-\frac{R}{R_c}\right).
	\label{SigmaProf}
\end{equation}
We then truncate this profile at different ``disc outer radii'' $R=R_d$ to yield a corresponding disc outer surface density $\Sigma_d$. The initial disc mass is related to $\Sigma_c$ via
\begin{equation}
	\Sigma_c = \frac{M_g}{2\pi R_c^2\left\{1-\left[\exp(-R_d/R_c)\right]\right\}}. 
	\label{equn:discmass}
\end{equation}
Interior to $R_d$, this disc portion of the calculation grid acts as a boundary condition to the flow. Outside of $R_d$ the medium dynamically evolves until it reaches as steady state. Note that these disc profiles have no bearing on our later viscous evolutionary calculations, they are simply used to provide representative surface densities at different disc radii from which we compute  a mass loss rate grid. For each model the grid extends out to 533\,AU and is adaptive -- refining to higher resolution automatically in regions which require it. The maximum effective number of cells is 2048, which yields a resolution of approximately 0.26\,AU.

\begin{figure*}
	\includegraphics[width=8.805cm]{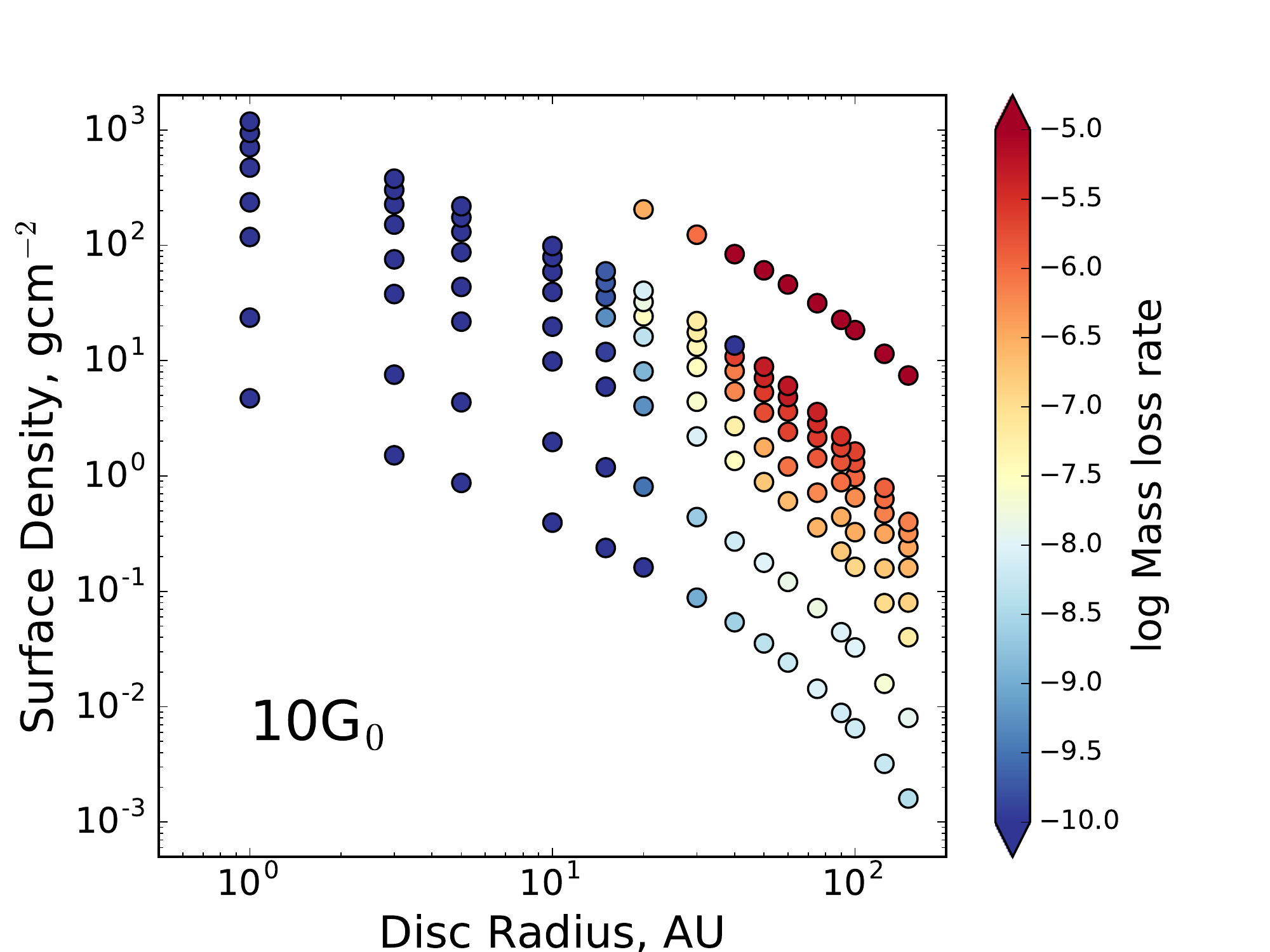}
	\includegraphics[width=8.805cm]{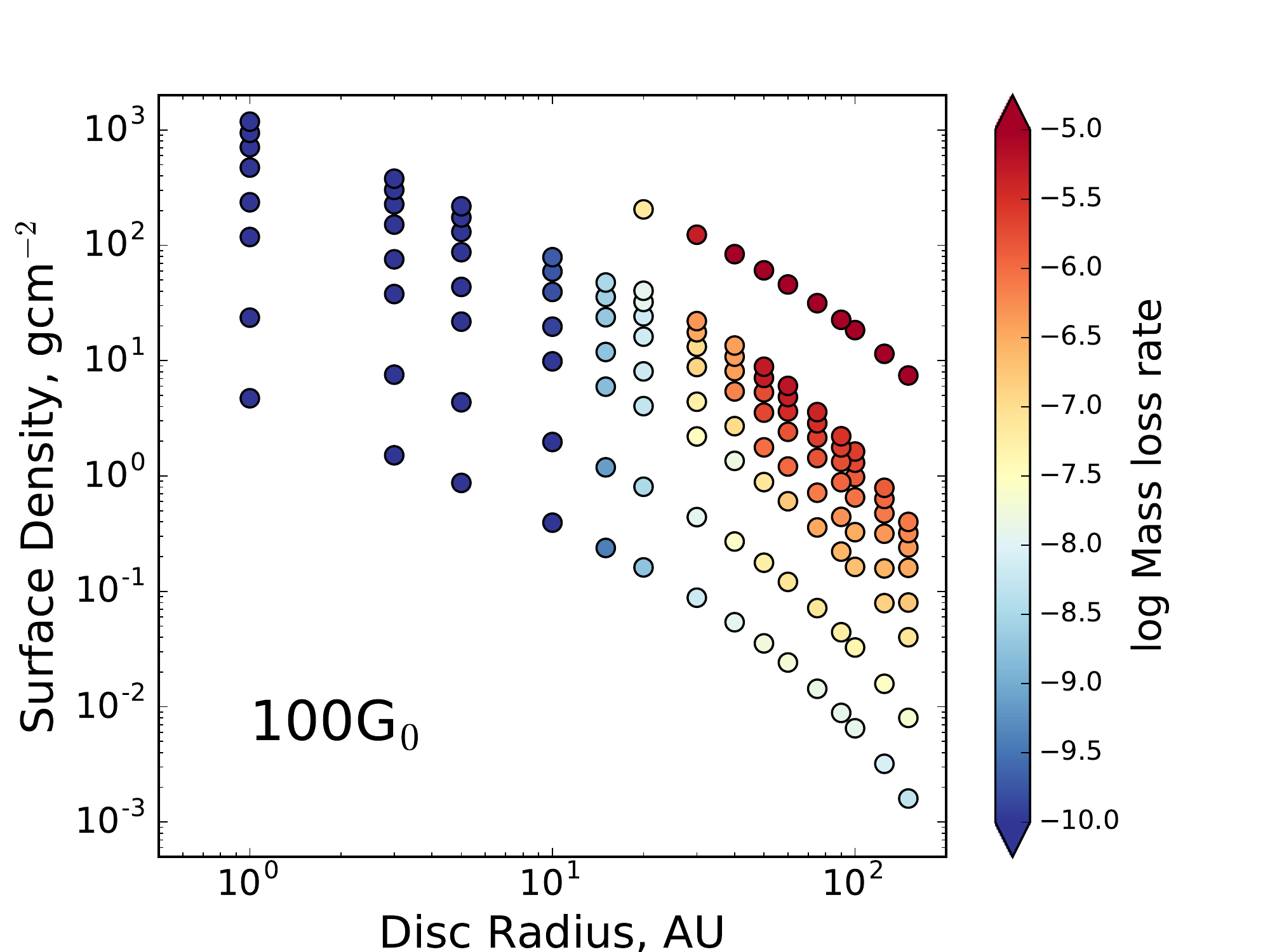}
	\includegraphics[width=8.805cm]{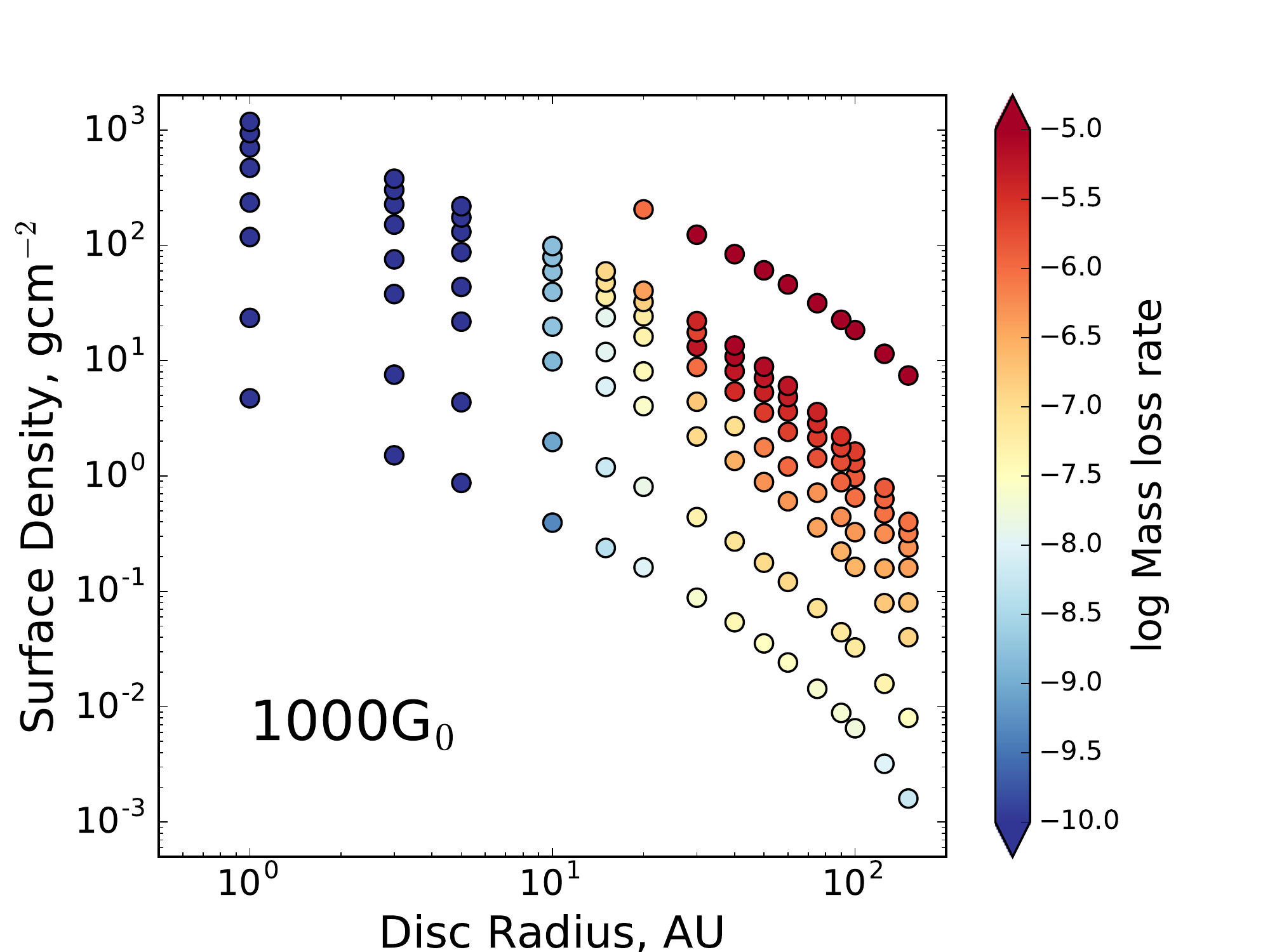}
	\includegraphics[width=8.805cm]{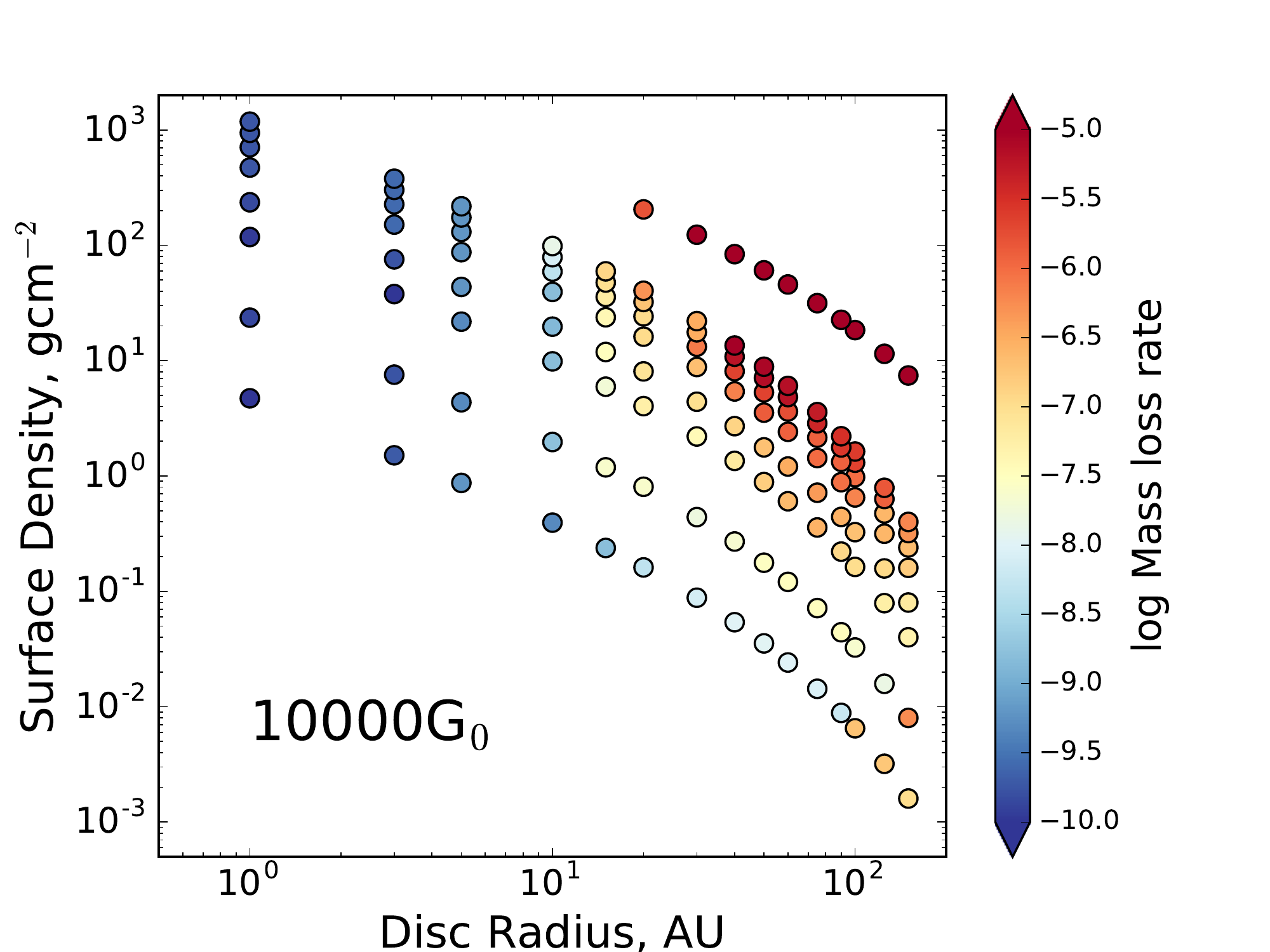}
	\caption{A summary of our grid of mass loss rates as a function of disc outer radius and surface density. Panels are for an incident radiation field of 10, $10^2$, $10^3$ and $10^4$\,G$_0$ from left to right, top to bottom. Each point is colour coded corresponding to the mass loss rate.  }
	\label{MdotGrid}
\end{figure*}

 We make one modification compared to our prior photoevaporation calculations \citep{2016MNRAS.463.3616H, 2017MNRAS.468L.108H}  which is to impose a limit on where the photodissociation region (PDR) microphysics is applicable in this 1D framework. Although likely not a problem in higher dimensions, in 1D models high density, highly optically thick regions can lead to spurious heating in PDR codes as the escape probability of cooling photons tends to zero. Therefore, interior to the point in the flow with an extinction greater than 10 and a gas local number density exceeding $10^{8}$\,cm$^{-3}$ we assume that the temperature of the flow is constant until heating from the star dominates.  This is the approach used by \cite{2016MNRAS.457.3593F} in the high extinction/density parts of their flow.  The temperature set by the star is assumed to be of the form $T_*=50\textrm{K}(R/\textrm{AU})^{-1/2}$. In our photoevaporation calculations at any radius we use the maximum of the temperature set by the star and that computed by \textsc{torus-3dpdr}.

Overall we ran around 550 steady state mass loss rate calculations, spanning a range of radii and surface densities, and for UV fields of $10$, $10^2$, $10^3$ and $10^4$\,G$_0$ \citep[the median value for stellar systems is of order $10^3$\,G$_0$, ][]{2008ApJ...675.1361F}.  Our resulting mass loss rate grids are summarised in Figure \ref{MdotGrid}. The mass loss rate initially increases as we move towards smaller radii, since there is more material at the base of the flow to resupply the wind. However eventually the disc becomes dense and compact enough that moving to smaller radii reduces the mass loss rate. At some radius, which is smaller for higher radiation field strengths, the mass loss rate drops to a very low value $\leq10^{-10}$\,M$_{\odot}$\,yr$^{-1}$. Typically increasing the disc mass leads to slightly higher mass loss rates (over the range of disc parameters considered here), but this is by far a secondary effect compared to the mass loss rate sensitivity to disc outer radius. This grid provides us with the mass loss rates for a sensible range of disc sizes and outer surface densities to interpolate over in our viscous evolutionary models.

\subsection{Viscous evolution}

\subsubsection{Evolutionary model method}
We compute the evolution of our discs using a modified version of the code developed by \cite{2007MNRAS.376.1350C}, {which follows the viscous evolution, accretion and external photoevaporation of the disc}. For each model we define an initial surface density profile of the form given by equation \ref{SigmaProf}  and initial mass accretion rate. {The initial disc structure and accretion rate set the accretion and viscous timescales, and hence the viscosity $\nu$, which we assume scales linearly with $R$. If the surface density profile in equation \ref{SigmaProf}  is interpreted in terms of a viscous steady state then for a temperature profile with $T \propto R^{-1/2}$ the Shakura Sunyaev $\alpha$ parameter is independent of radius and may be estimated by}
\begin{multline}
	\alpha =  \frac{R_c^2}{3}\frac{\dot{M}_a}{M_g}\frac{\Omega}{c_s^2} \\
	\approx 3\times10^{-3} \left(\frac{R_c}{50\textrm{AU}}\right)^{2} \left(\frac{\dot{M}_a}{6.4\times10^{-10}M_\odot\,\textrm{yr}^{-1}}\right) \left(\frac{M_g}{0.01M_*}\right)^{-1}\times \\ 
	\left(\frac{c_s}{\textrm{km}\,\textrm{s}^{-1}}\right)^{-1}\left(\frac{H}{\textrm{AU}}\right)^{-1}
	\label{equn:alpha}
\end{multline}
{where $c_s$, $\Omega$, $H$ and $\dot{M}_a$ are the sound speed, angular velocity, scale height and mass accretion rate. Both $\nu$ and $\alpha$ are constant in time.  The disc then evolves under the combined action of viscous evolution, accretion and external photoevaporation. We do not account for any change in the disc evolution from the dead zone into the inner disc (i.e. we ignore magnetohydrodynamic effects). }

The initial conditions of the discs in our evolutionary models are set up using equations \ref{SigmaProf}, \ref{equn:discmass}. These calculations use  330 cells logarithmically spaced between 0.5 and 300\,AU.  Our grid of models sample $R_c$ from 10 to 100\,AU and initial disc mass from 3 to 50\,per cent of the stellar. We assume an initial accretion rate onto the star of $ 6.4\times10^{-10}$\,M$_{\odot}$\,yr$^{-1}$, which we obtain using $\dot{M}_a\propto M_{}^{2}$ scaling down from $10^{-7}$\,M$_{\odot}$\,yr$^{-1}$ for a disc about a 1\,M$_{\odot}$ star \citep{2006ApJ...639L..83A}. This initial accretion rate is consistent with recent observed accretion rates for low mass objects observed by \cite{2014A&A...561A...2A}, \cite{2016A&A...585A.136M}.  This initial accretion rate \textit{will} evolve over time if the disc viscously evolves (i.e. in mass/extent) in such a way that modifies it. {\cite{2013ApJ...774....9A} demonstrated that higher viscosity leads to more effective photoevaporation. From equation \ref{equn:alpha} we have a canonical $\alpha\approx10^{-3}$ for discs with mass that is 1\,per cent the stellar, and $\alpha$ will be lower for more compact and/or higher mass discs. Over the vast majority of our parameter space we therefore expect a smaller $\alpha$, and hence less effective photoevaporation, making our models conservative. }

\begin{figure}
	\hspace{-0.6cm}
	\includegraphics[width=9cm]{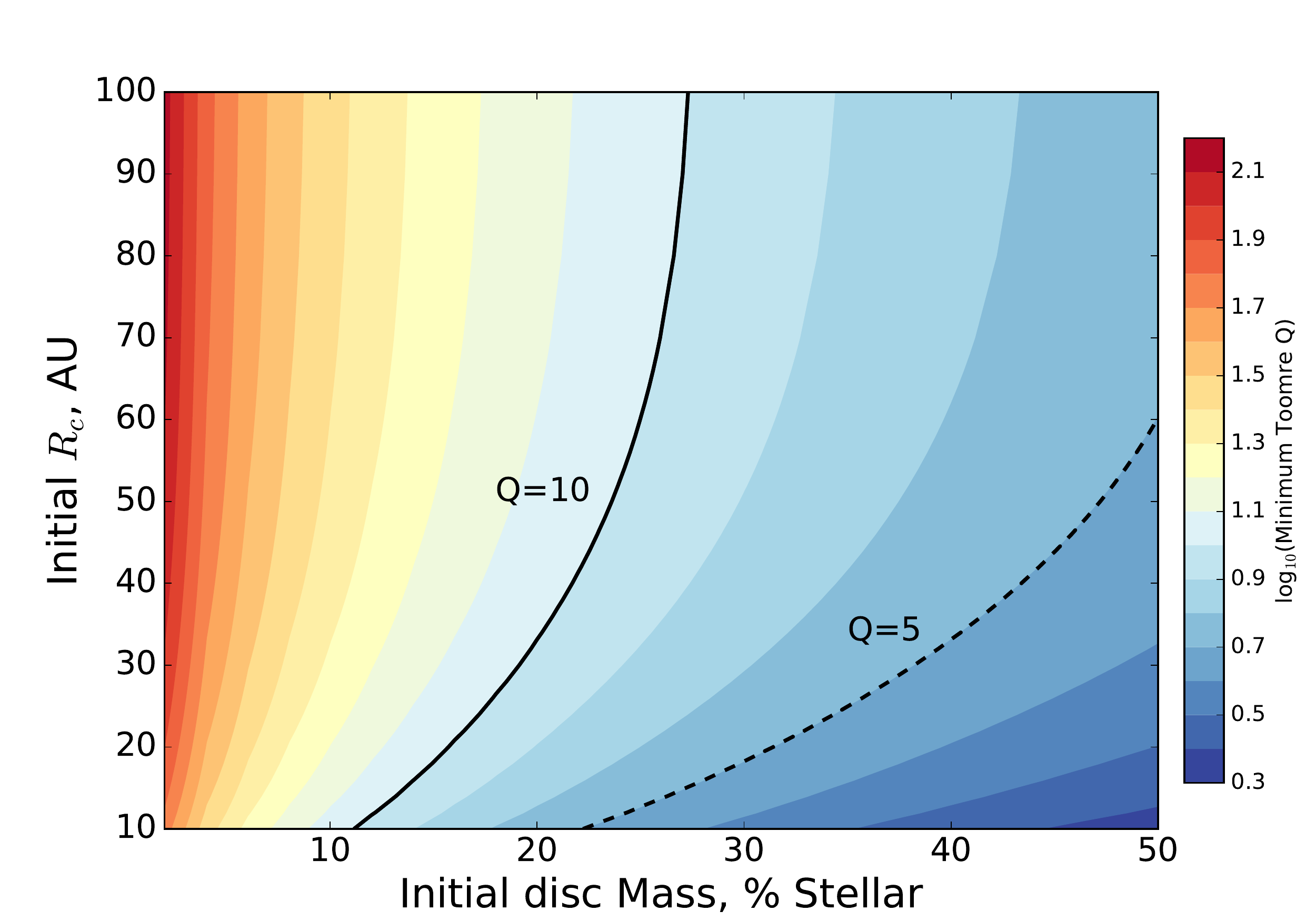}
	\vspace{0cm}
	\caption{The minimum value of the Toomre Q parameter in our parameter space of disc properties. Our discs are all expected to be gravitationally stable ($Q>1$). The contours denote the $Q=10$ (solid) and $Q=5$ (dashed) boundaries.}
	\label{minToomreQ}
\end{figure}

We checked the \cite{1964ApJ...139.1217T} stability parameter
\begin{equation}
	Q = \frac{c_s\Omega}{\pi G \Sigma}
	\label{equn:ToomreQ}
\end{equation}
for each of the models that we will discuss, finding that they are all gravitationally stable ($Q>1$) at all radii in the disc.  To illustrate this, in Figure \ref{minToomreQ} we plot the minimum $Q$ value over the entire disc for each model in our parameter space. {Our approach would not have been suitable for modelling gravitationally unstable discs, which would require 3D photochemical-dynamical calculations that are currently impossibly comptuationally expensive.}

The surface density at the outer edge of a disc for a given UV field strength is  interpolated over the photoevaporative mass loss grid of section \ref{mdotcalc} to yield an evaporative mass loss rate. The evolution of the extent of the disc depends on the relative values of the flux of material through viscous spreading (driven by accretion) and mass lost in the wind. If the photoevaporative wind is stronger than the mass flux through viscous spreading, then the disc will shrink over time, and vice versa \citep[e.g.][]{2007MNRAS.376.1350C, 2017MNRAS.468.1631R}.

\subsubsection{Stripping of dust in the photoevaporative wind}
\label{sec:dustStrip}
As the disc is evaporated and shrinks (or expands) it will deplete the disc of dust and possibly influence planet formation. Not all dust at the disc outer edge is entrained in the wind as the gas is stripped. \cite{2016MNRAS.457.3593F} showed that the maximum grain size entrained is roughly given by 
\begin{equation}
	a_{\textrm{entr}} \approx \frac{v_{th}}{GM_*}\frac{1}{4\pi \mathcal{F} \rho_g}\dot{M}_w
	\label{equn:grainEntrainMax}
\end{equation}
where ${\rho_g}$ is the density of the grains themselves, assumed to be 1\,g\,cm$^{-3}$, and $v_{th}=\sqrt{8k_BT/\pi/\mu/m_H}$. 
The dust mass loss rate for a distribution with maximum grain size $a_{max}$ is hence 
\begin{equation}
	\dot{M}_{dw} = \delta\dot{M}_w\left(\frac{a_{\textrm{entr}}}{a_{\textrm{max}}}\right)^{1/2}
	\label{equn:dustmdot}
\end{equation}
for dust-to-gas mass ratio $\delta$ and power law of the grain size distribution $q=3.5$ \citep[a][grain distribution]{1977ApJ...217..425M}. Therefore over time, as grains grow, less dust is entrained in the wind. 

We assume that initially the minimum grain size throughout the disc is $0.01\,\mu$m. This grows over time at each radius in the disc according to 
\begin{equation}
	\frac{\textrm{d}a}{\textrm{d}t} = \frac{a}{\tau}
	\label{equn:amaxOverTime}
\end{equation}
\citep{2012A&A...539A.148B} where $\tau = 1 /\delta \Omega_k$ and $\Omega_k=(GM_*/R^3)^{1/2}$. Since we are here concerned with the growth of grains prior to their attaining a size where they drift relative to the gas, we employ a fixed value of $\delta=10^{-2}$ in equation \ref{equn:dustmdot} and thereby also neglect the enhancement of $\delta$ in the outermost disc due to the fact that larger grains are also not entrained in the wind.

The maximum grain size could be set either by fragmentation, or by radial drift. Radial drift limits the grain size once the Stokes number is $\sim0.1$ \citep{2017MNRAS.469.3994B} hence 
\begin{equation}
	a_{d}=0.1\frac{\Sigma}{\rho_g}
	\label{equn:adequn}
\end{equation}
is the drift limited maximum grain size. Similarly, the fragmentation limited maximum grain size is 
\begin{equation}
a_{f}= 0.37\frac{1}{3\alpha}\frac{v_f^2}{c_s^2}\frac{\Sigma}{\rho_g}
\end{equation}
\citep{2017MNRAS.469.3994B} where $v_f$ is the fragmentation velocity, which we assume to be $10$\,m\,s$^{-1}$ \citep{2015ApJ...798...34G}. If we then take the ratio $a_f/a_d$, with $\alpha$ from equation \ref{equn:alpha}
\begin{equation}
	\frac{a_f}{a_d} = 3.7 v_f^2 f M_*^{1/2}R_c^{-1/2}G^{-1/2}\dot{M}_a^{-1}
\end{equation}
and substitute in the Trappist-1 stellar mass, our assumed initial accretion rate and $v_f=10$\,m\,s$^{-1}$ then
\begin{equation}
	\frac{a_f}{a_d} = 1150 f \left(\frac{R_c}{\textrm{AU}}\right)^{-1/2}
\end{equation}
where $f$ is the ratio of the initial disc to stellar masses $M_d/M_*$. For the parameter space considered here, we therefore expect to have $a_f > a_d$ except at very small radii and hence drift will dominate in setting the maximum grain size in the outer parts of the disc from which the wind is launched. 

We note that once the maximum grain size first
reaches $a_d$ at the outer edge of the disc we do not explicitly calculate
the dust evolution thereafter. We however assume that after this point
the dust can drift to the water snow-line and form planets efficiently, which then reduces the dust content of the material that accretes on to the star.
We conservatively assume that {\it no} dust is accreted on to the star
once the grain size at the outer edge attains $a_d$. Therefore, in what follows we assume that once the Stokes number at the disc outer edge attains a value of 0.1, widespread radial drift of grains into the water snow line shuts off further accretion of dust (though gas continues to accrete).





For simplicity, we only follow the total mass of grains entrained in the wind (and by extension the mass remaining in the disc),  this would be an extremely conservative upper limit on the total planetary mass that could be produced.  Not all dust remaining will drift in to the water snow line \citep[][permit 50\,per cent to do so]{2017A&A...604A...1O}. Furthermore once at the water snow line the pebble accretion efficiency is also not 100\,per cent, with \cite{2017A&A...604A...1O} using a value of 25\,per cent. Additional pressure bumps might also strand dust at larger radii \citep{2012A&A...538A.114P}. We will proceed only accounting for dust lost in the wind, but will consider additional factors such as the pebble accretion efficiency when discussing the potential for planet formation from our discs in section \ref{sec:where}. 

In principle if the dust and gas masses are comparable then feedback from the dust dynamics onto the gas could become important, however we do not account for this. In what follows we will show that once the global dust to gas ratio approaches unity the gas is rapidly removed through photoevaporation anyway, so we don't expect this to have a significant impact on our models. 

In appendix \ref{sec:appendix1} we show that equations \ref{equn:grainEntrainMax}-\ref{equn:amaxOverTime}, combined with our photoevaporative mass loss rates, can be used to get a semi-analytic estimate of the regions of a disc that are resilient to photoevaporation over time.

\subsection{Quick recap of our methodology}
Since there are a lot of components to our method, we quickly recap our general approach. We are running a grid of 1D viscous evolutionary models of protoplanetary discs, including the effects of accretion and photoevaporation. Prior to these viscous evolutionary calculations, we have computed a grid of photoevaporative mass loss rates as a function of disc size and outer surface density, which we then interpolate over to give the mass loss rate in the evolutionary models. We keep track of the mass of dust that remains in the disc and is potentially available for planet formation, taking account of mass loss on to the star due to accretion (before grains in the outer disc reach $St=0.1$) and entrainment in the wind.

\section{Results and discussion}

\subsection{Detailed evolution of illustrative discs}

Our evolutionary models contain a lot of information, such as the disc gas/dust masses over time, the spatial extent of the disc, the mass loss rates and so on. Therefore, before discussing the global properties of our large grid of evolutionary models, we begin by studying a handful of models in more detail to aid in the conceptual interpretation of the macroscopic grid properties. We consider two illustrative discs with initial masses of 15 and 35\,per cent the stellar respectively and $R_c=50$\,AU in both cases (equations \ref{SigmaProf}, \ref{equn:discmass}). For these two discs, we consider radiation environments of $10$, $10^2$, $10^3$ and $10^4$\,G$_0$

\label{sec:focussedEvo}

\begin{figure}
	\includegraphics[width=9cm]{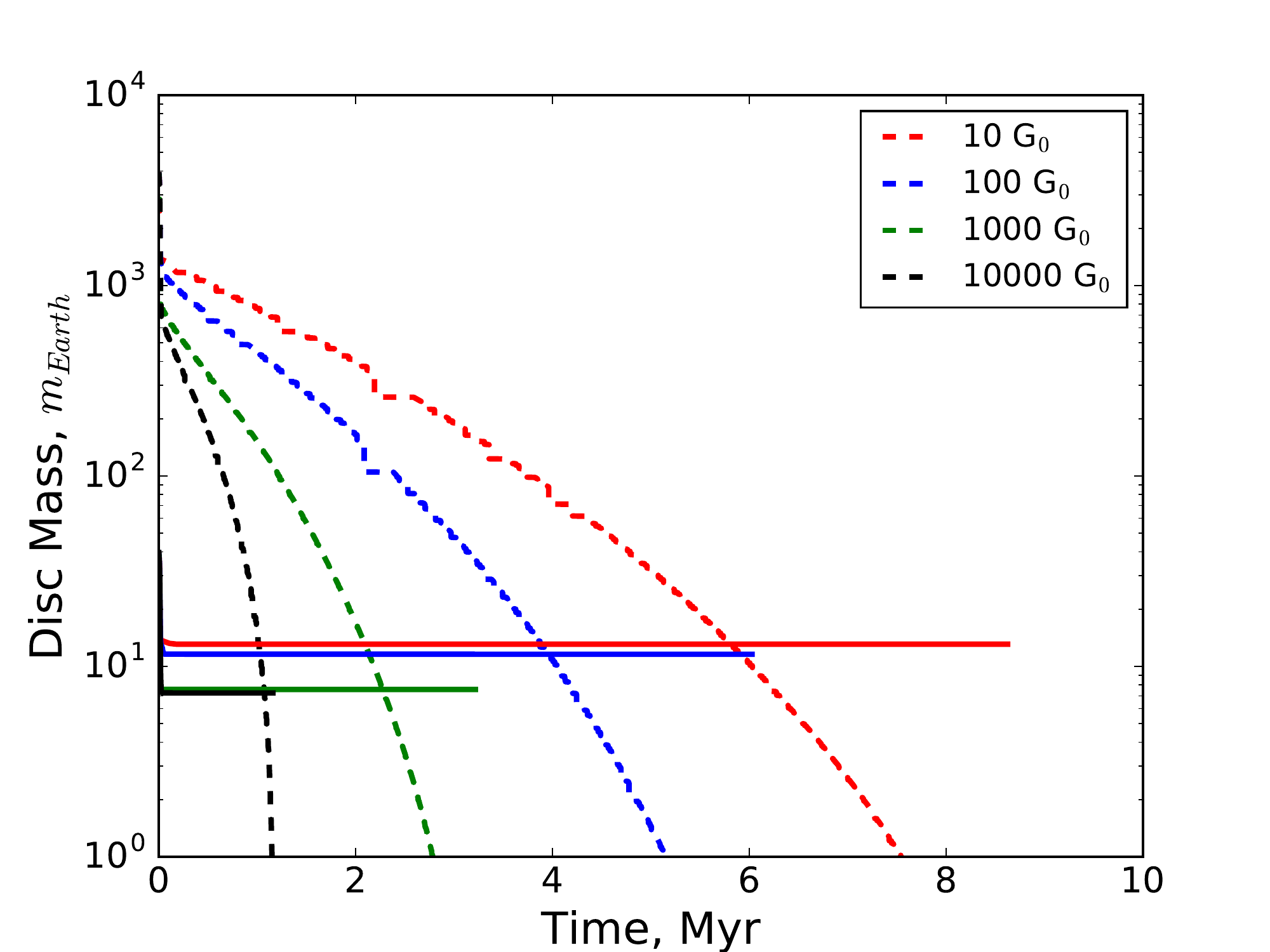}

	\includegraphics[width=9cm]{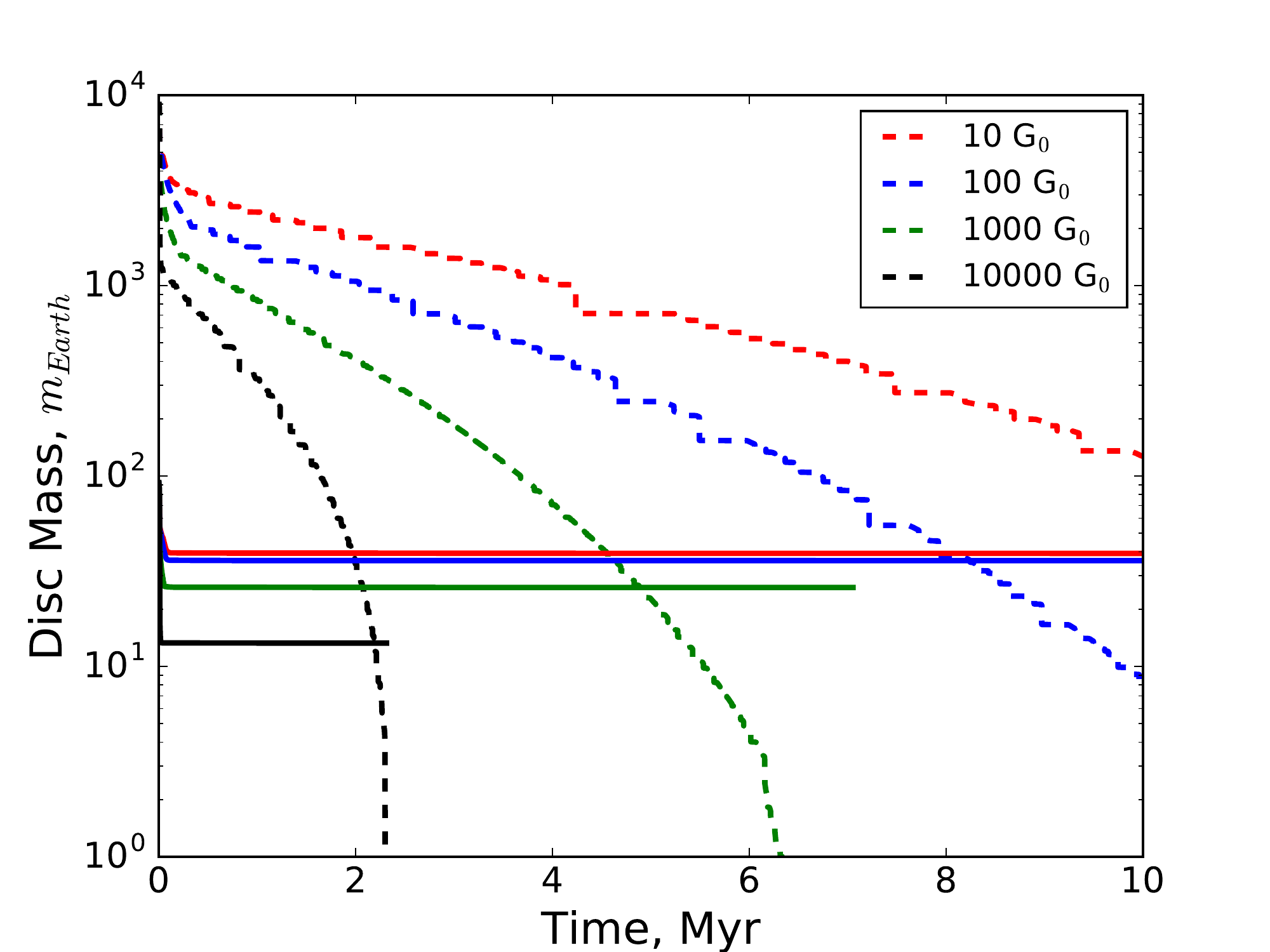}
	\caption{The evolution of the gas (dashed line) and solids (dust and planets, solid line) mass in our illustrative discs as a function of time. Different coloured lines correspond to different irradiating UV field strengths. The upper and lower panels are for initial disc masses of 15 and 35\,per cent the stellar respectively (both discs have $R_c=50$\,AU).  }
	\label{fig:detailedMasses}
\end{figure}

\begin{figure}
	\includegraphics[width=9cm]{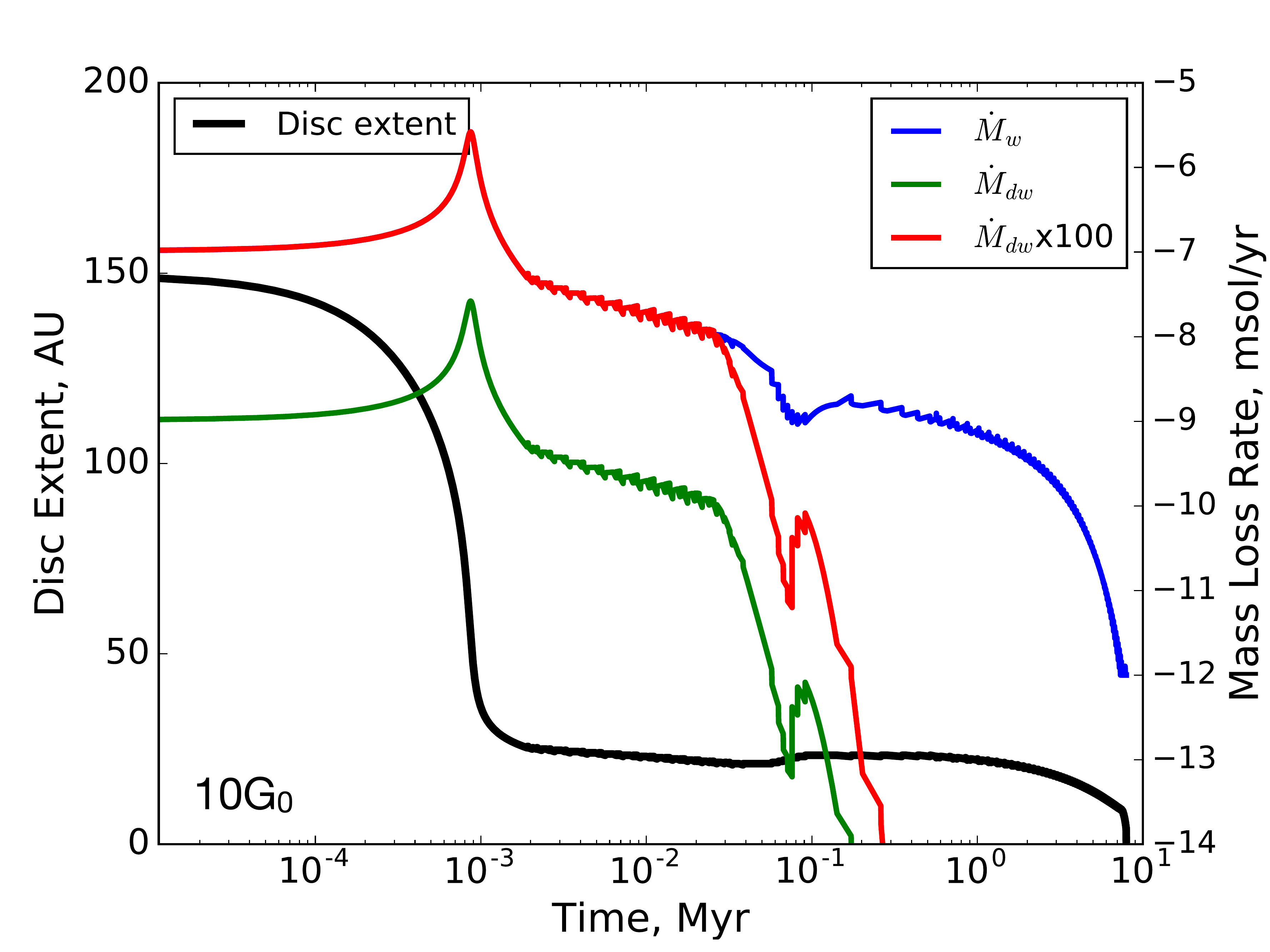}

	\includegraphics[width=9cm]{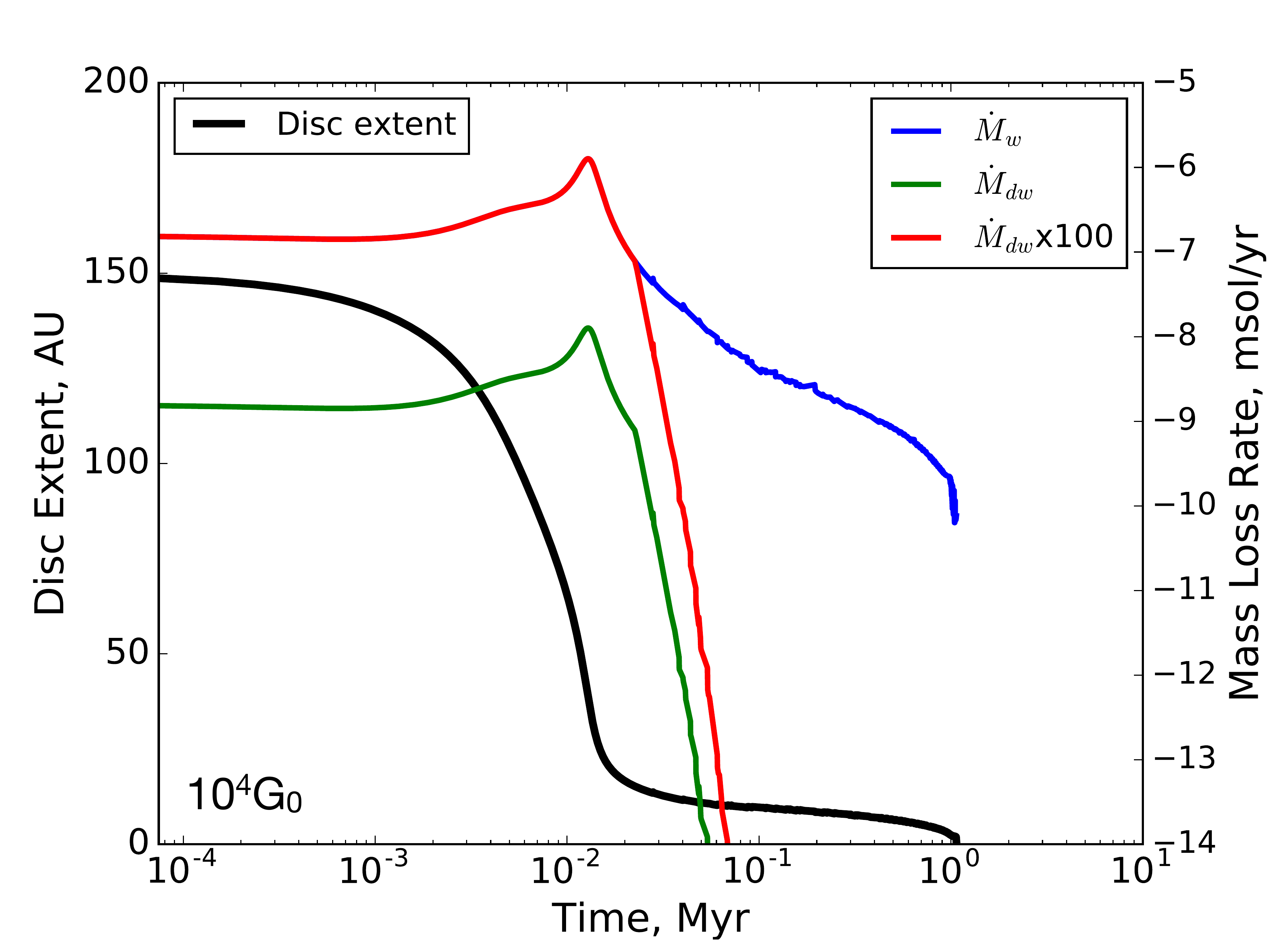}
	\caption{The disc extent (black line) and mass loss rates (due to accretion and photoevaporation of gas and dust) as a function of time for a model with initial disc mass 15\,per cent the stellar, $R_c=50$\,AU and incident UV fields of 10 and 10000\,G$_0$ in the upper and lower panels respectively.  The blue and green are the gas and dust mass loss rates in the wind respectively. The red is the dust mass loss rate multiplied by the initial gas-to-dust mass ratio. So where the red is not coincident with the blue, grain growth is hindering the entrainment of grains in the wind.  }
	\label{fig:detailedExtentMdots}
\end{figure}

Figure \ref{fig:detailedMasses} shows the evolution of the disc mass over time for these illustrative discs. The dashed lines are the gas mass and the solid line is the mass in solids (planets and dust). Different coloured lines denote different radiation field strengths. To assist in interpretation of these mass evolutionary plots we also include the spatial extent and mass loss rates for two of the models in Figure \ref{fig:detailedExtentMdots}.

\begin{figure}
	\includegraphics[width=9cm]{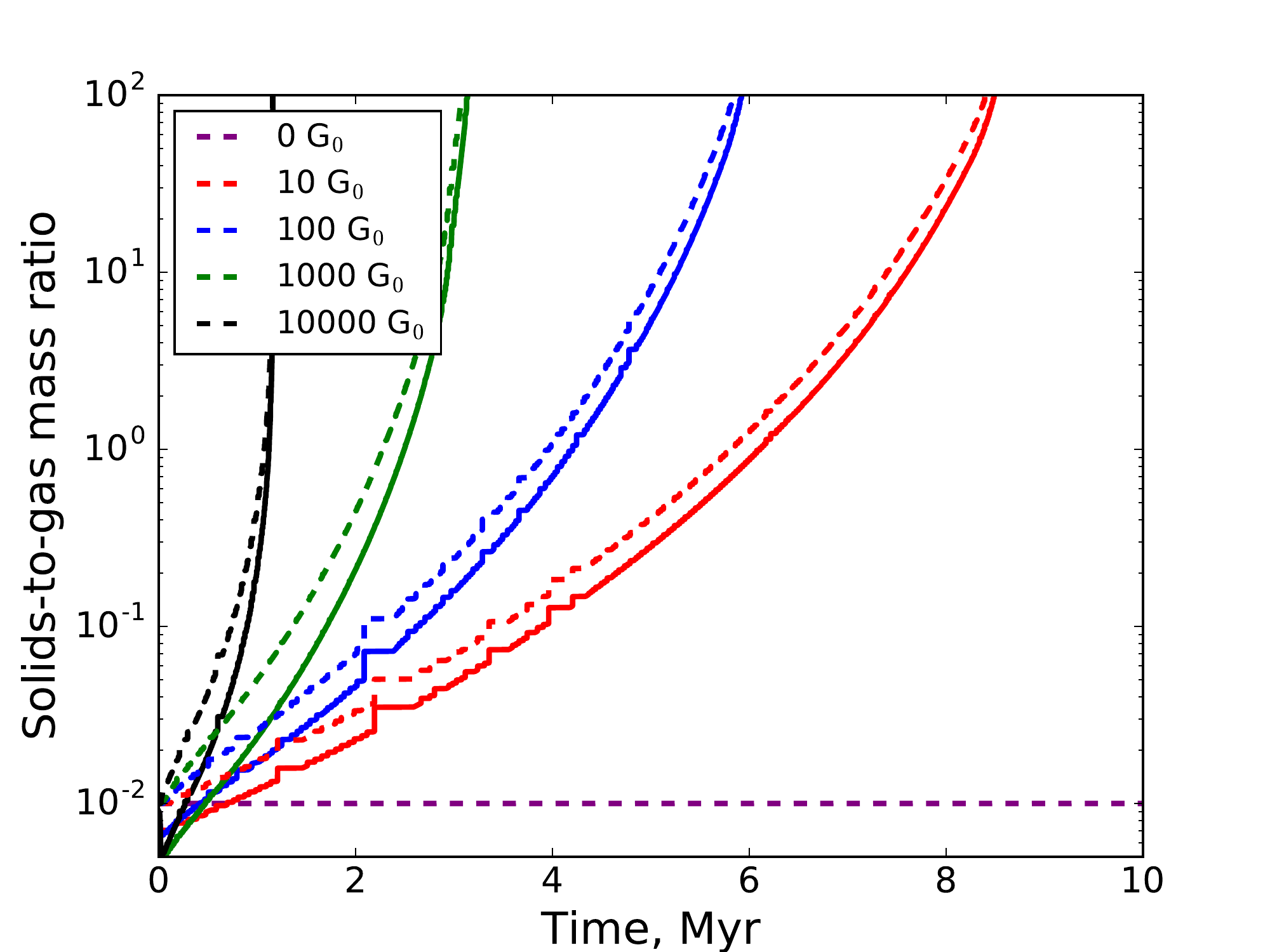}

	\includegraphics[width=9cm]{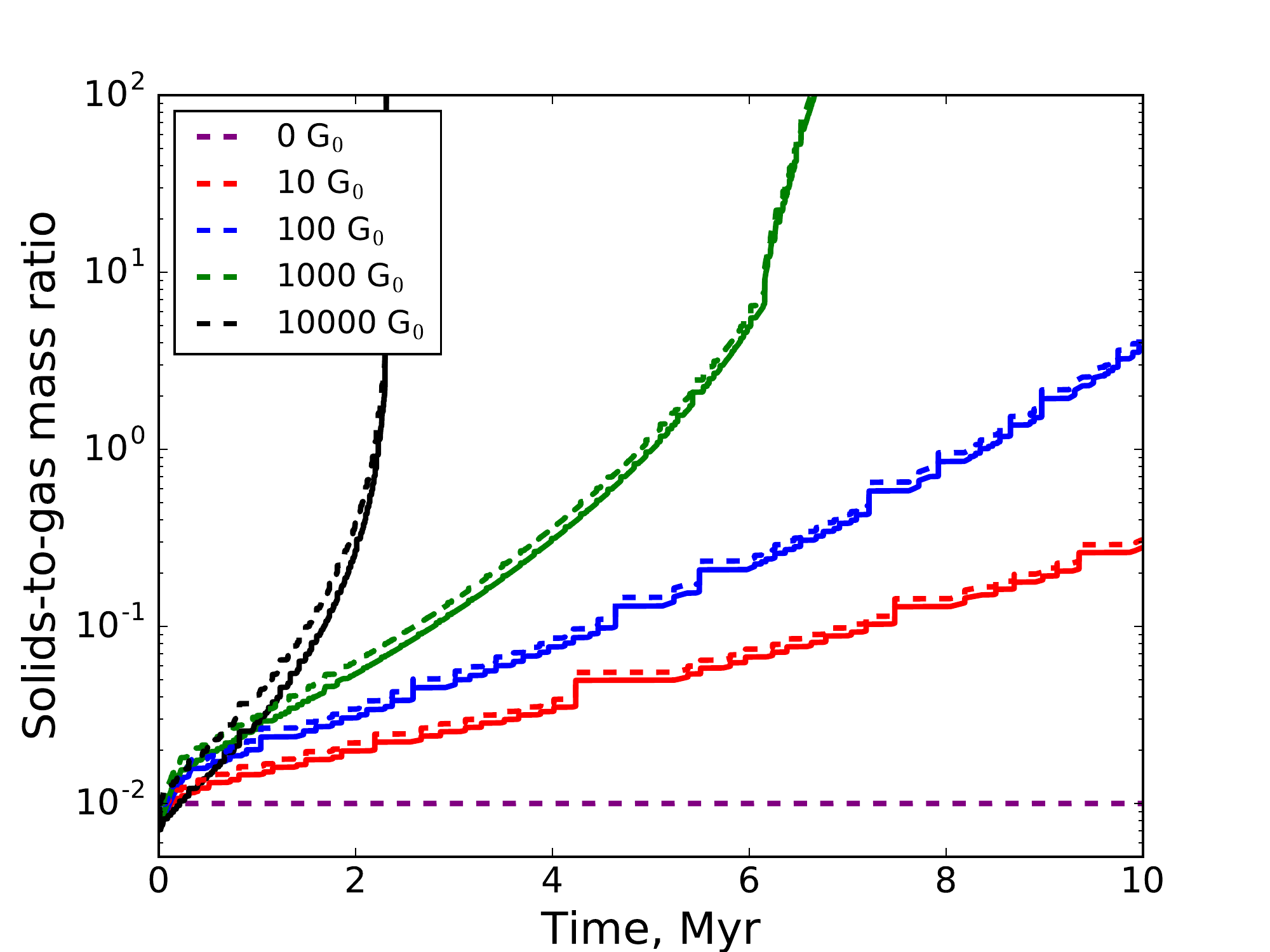}
	\caption{The evolution of the global solids-to-gas mass ratio in our illustrative discs. The dashed line is the mass of the mass in dust plus planets and the solid line is neglects 4\,$M_{\oplus}$ at all times, which is the mass in planets detected about Trappist-1. Different coloured lines correspond to different irradiating UV field strengths. The upper and lower panels are for initial disc masses of 15 and 35\,per cent the stellar respectively (both discs have $R_c=50$\,AU). {Note that dynamical feedback from dust onto the gas is not included in our models.} }
	\label{fig:detailedDustToGas}
\end{figure}

Initially there is a period of photoevaporation-induced rapid truncation of the disc in which a substantial fraction of gas and dust mass is lost (e.g. quite readily 50--80\,per cent of the initial mass, depending on the UV field and initial disc parameters). The disc then stalls (because the flux in material through viscous spreading balances the mass loss rate in the wind) and undergoes slower photoevaporation for the bulk of its lifetime until the gas is depleted. During the early strong depletion phase the dust is small enough to also be readily stripped from the disc.  However, after only a few tens of kyr grain growth means that a significant fraction of the remaining dust population is too large to be entrained in the photoevaporative wind. This is illustrated by the mass loss rates in Figure \ref{fig:detailedExtentMdots}. Where the blue and red lines coincide in that Figure, the dust is perfectly entrained in the flow.  However by about 30\,kyr  the dust mass loss rate drops to a lower value as grain growth occurs. By 0.1\,Myr the dust mass loss rate is negligible in both cases.

The effect of the dust growing to a size at which it is not entrained in the photoevaporative wind is that although the gas disc steadily depletes, the dust disc is pretty much completely resilient to photoevaporation after this $\sim30$\,kyr (at most 0.1\,Myr) timescale.  The dust mass hence plateaus after this time (see Figure \ref{fig:detailedMasses}). This kind of behaviour is a general feature of our models, and when we consider the entire grid it is the level of this plateau in dust mass that will be important for constraining where Trappist-1 planet formation is possible.

\begin{figure}
		\includegraphics[width=9cm]{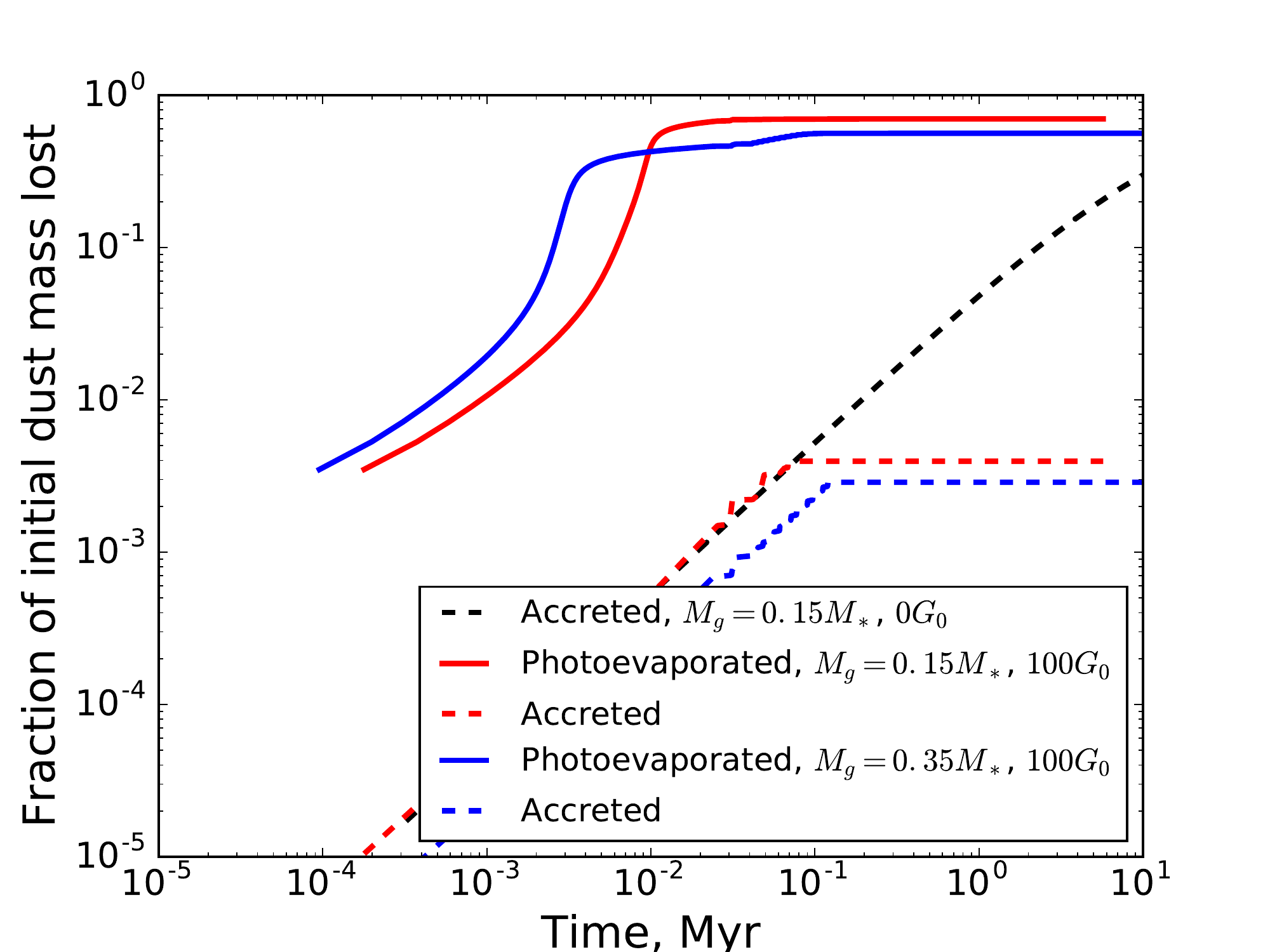}
	\caption{The total dust mass lost dust to photoevaporation (red) and accretion (blue) over time for our illustrative disc calculations in a 100\,G$_0$ radiation field. When the photoevaporation mass loss plateaus it is because grain growth means little dust is entrained in the flow. When the accretion mass loss plateaus there is assumed to be efficient accumulation of solids at the water snow line which quenches further accretion of solids onto the star. }
	\label{fig:totalMassLost}
\end{figure}

An interesting feature of our models is that we can follow the global solids-to-gas mass ratio over time. We refer to the ``solids'' rather than dust, because some fraction of this mass may end up in planets or other bodies that would not be accounted for in a dust-to-gas mass measurement. In Figure \ref{fig:detailedDustToGas} we plot the solids-to-gas mass ratio for our illustrative discs over time. The dashed lines account for the total dust plus planetary mass and the solid lines subtract 4\,$M_{\oplus}$ (the observed planetary mass in Trappist-1) of material. Subtracting the observed planetary mass doesn't have a significant effect on the global dust-to-gas ratio, which rapidly deviates from the canonical ISM value of $10^{-2}$ in all cases, irrespective of the UV field (over the range of parameters we consider). 

A population of such discs of random age would therefore be likely to have an elevated dust-to-gas mass ratio unless the vast majority of their residual
solids (typically tens of Earth masses, Figure 4) is contained in unobservably large rocky bodies. Evidence for enhanced \textit{dust} to gas mass ratios have been claimed in a number of systems, though currently these are typically all in a higher mass regime  \citep[e.g.][]{2010A&A...518L.124M, 2012ApJ...747..136I, doi:10.1093/pasj/psv098, 2016ApJ...816...25P, 2016MNRAS.461..385B, 2016ApJ...828...46A, 2017ApJ...844...99L}. Our models suggest that external photoevaporation significantly alters the global dust-to-gas mass ratio of protoplanetary discs by more effectively stripping the gas than the dust, at least about low mass stars like Trappist-1. Understanding the magnitude of this effect for the discs of higher mass stars will require future dedicated models. {We reiterate that dynamical feedback from grains onto the gas is not included in our models, which may become important at high dust-to-gas mass ratios.}

In the case of no irradiation in our models we never reach $St=0.1$ at the disc outer edge and so under our prescription, gas and grains accrete in constant proportion throughout the entirety of the disc evolution. The solids-to-gas mass ratio hence remains constant in this case.

\begin{figure*}
	\includegraphics[width=15cm]{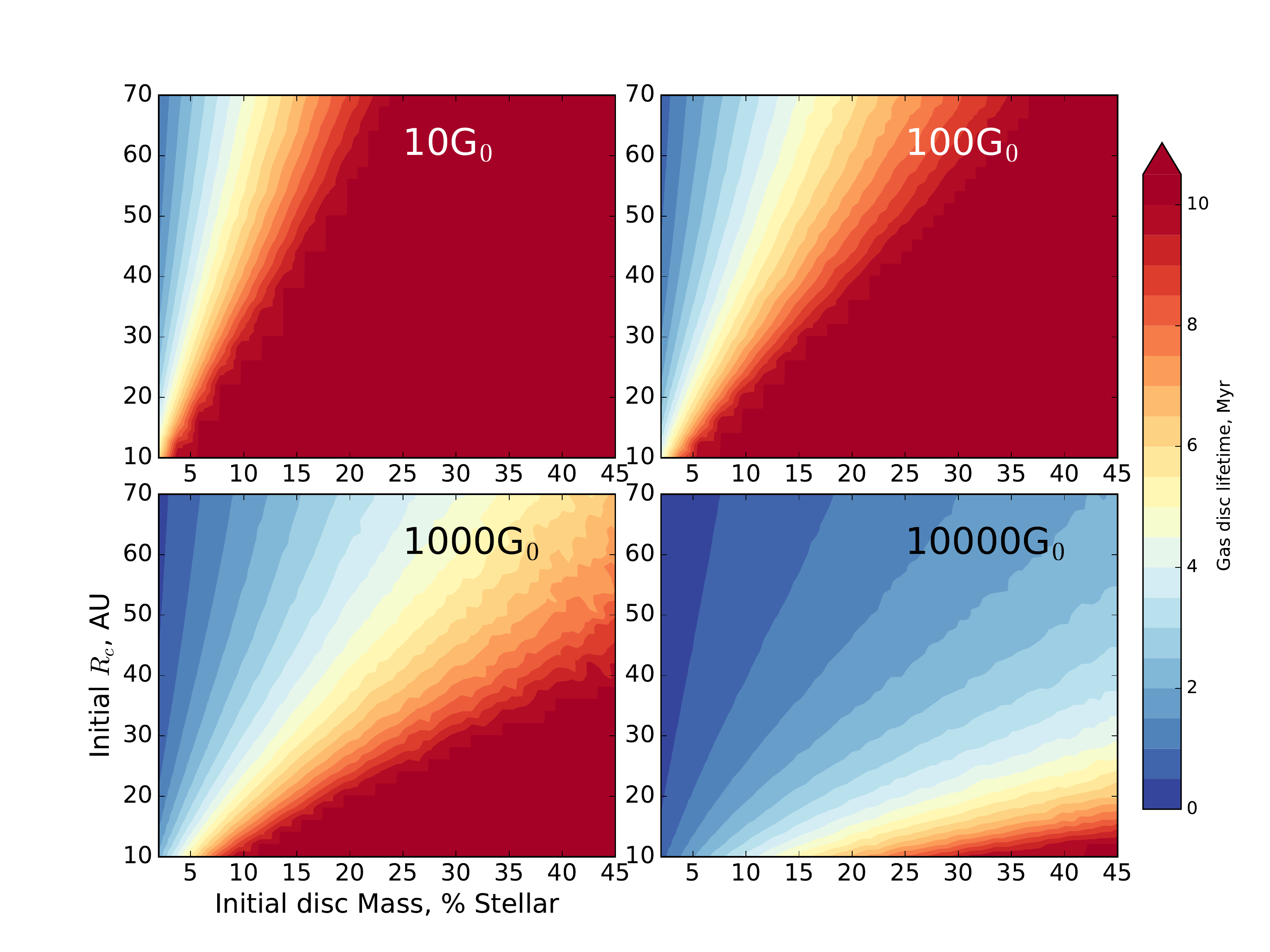}
	\caption{The time in our grid of models at which the gas is fully dispersed (which we also refer to as the debris disc onset time). The axes represent the range of initial disc masses and scaling radii, and the time at which the gas disc is dispersed is represented using the colour scale. }
	\label{discLifetime}
\end{figure*}

In Figure \ref{fig:totalMassLost} we show the total fraction of the dust mass that is photoevaporated and accreted over time for our example models in a 100\,$G_0$ environment (we will study the fraction of the dust mass accreted/photoevaporated over our entire grid in section \ref{sec:dustExtracted}). Effective accretion of dust is happening for a longer period of time  than stripping in the wind but at a rate that is orders of magnitude lower. The general result for all of our models is that the mass accreted in dust is negligible compared to that liberated in the wind. In Figure \ref{fig:totalMassLost}  the higher mass disc loses a larger fraction of its grains at a faster rate  early on, but plateaus at a lower value (since photoevaporation becomes less effective for a compact massive disc) meaning that a smaller fraction of the dust is depleted. Note however that about a factor of 2 more total (i.e. not fractional) dust mass is depleted in the case of the higher mass disc. In Figure \ref{fig:totalMassLost} we also include the accreted mass fraction for a model with no photoevaporation (black) and an initial disc mass 15\,per cent the stellar. By 10\,Myr around 30\,per cent of the initial dust mass is accreted in a $0$\,G$_0$ environment, compared to the 70\,per cent photoevaporated in the 100\,G$_0$ case (note that the fraction of initial dust mass photoevaporated is 67, 80 and 81\,per cent in the $10$, $10^3$ and $10^4$\,G$_0$ cases). So crucially, even in very weak UV environments depletion of the dust reservoir by photoevaporation dominates over that lost through accretion over the disc lifetime, even when the dust is allowed to accrete for the full 10\,Myr.

\begin{figure*}
	\includegraphics[width=15cm]{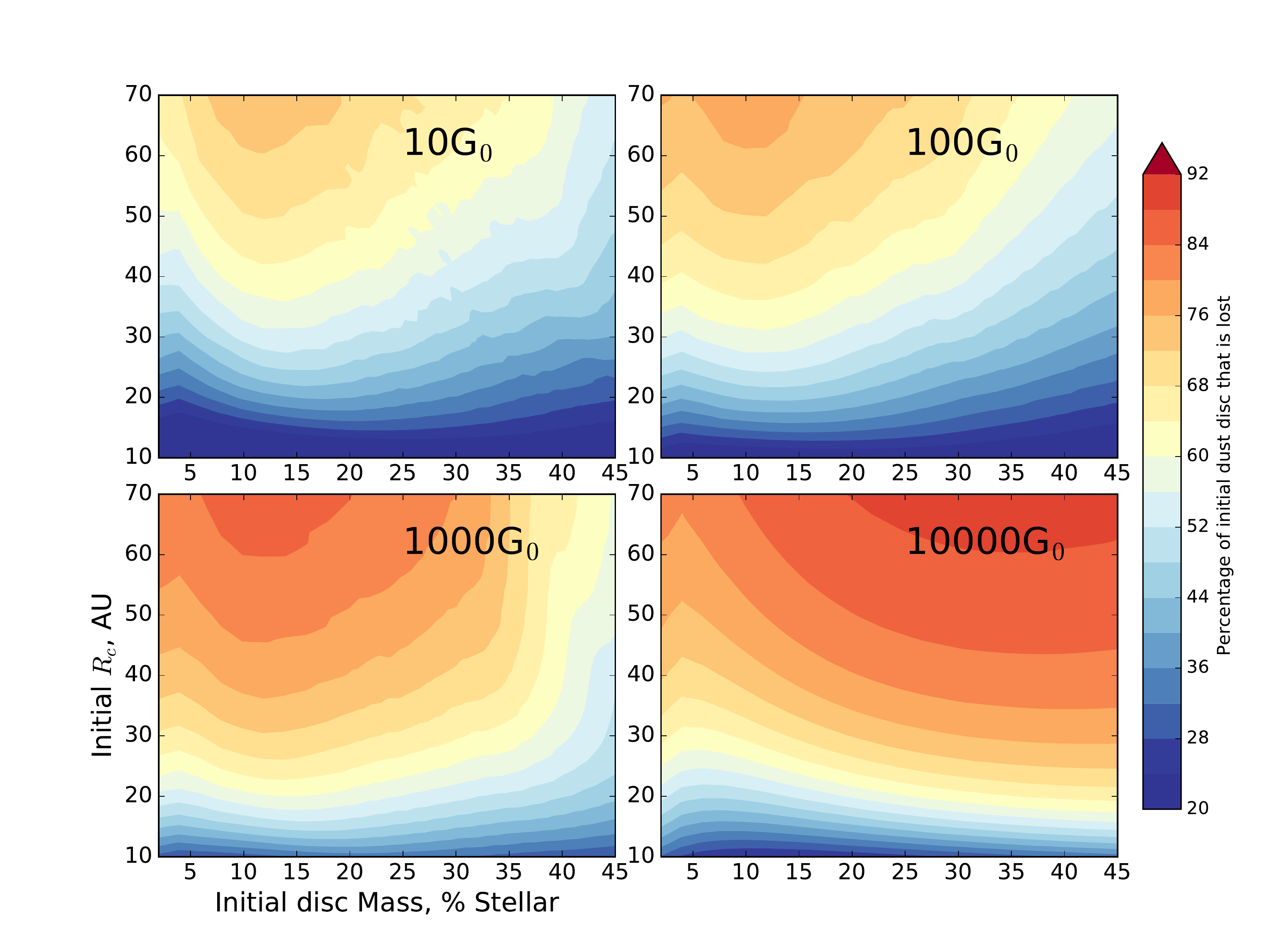}
	\caption{The perecentage of the initial mass in dust that is lost in our grid of evolutionary models.  }
	\label{lostDustFraction}
\end{figure*}

\subsection{The protoplanetary disc lifetime}
\label{sec:debrisdisc}

We now turn our attention to the behaviour of our large grid of evolutionary models. We first consider the time at which the gas mass drops to zero. Given that there is a dust reservoir remaining that is not entrained in the wind, and the final clearing of the disc happens rapidly, this time is also very close to that at which the dust mass dominates over the gas. This time is hence synonymous with the time at which an observer would infer that the disc is a debris disc. There is no formal definition of a debris disc \citep[see e.g.][]{2015Ap&SS.357..103W} so we use the gas disc lifetime and ``debris disc onset time'' interchangeably. 

This gas lifetime is plotted as a function of the initial disc mass and scaling radius $R_c$ in Figure \ref{discLifetime}. Note that we terminate our calculations after 10\,Myr, which is therefore the artificial upper limit on these plots.

The variation of the gas lifetime can be understood intuitively. More massive, compact discs are longest lived. This is because they have the largest mass reservoirs which are also most strongly bound to the parent star, weakening the power of external photoevaporation. Conversely extended (more weakly bound) lower mass discs are easily stripped of material even when the disc has shrunk (since they contain less mass). These discs are hence easily stripped of gas, whilst grain growth will permit a substantial fraction of the dust to survive without getting entrained in the wind.

Unsurprisingly, larger UV field strengths also lead to shorter gas disc lifetimes, owing to the more effective heating of the disc outer edge.  The most important impact of a higher UV field is that it is more capable of effectively heating and stripping material from the disc once it becomes compact ($\sim10-20$\,AU) and hence is able to completely remove the gas disc more quickly. In weaker UV environments the disc does shrink, but stalls at some radius and is long lived, as we discussed in section \ref{sec:focussedEvo}. 

For some spread in initial disc masses and radii, at a given time our models predict that we should expect to find mixtures of protoplanetary and debris discs, which is consistent with observations of similar aged young clusters \citep[e.g.][]{2006ApJ...638..897S, 2007ApJ...662.1067H}. We also expect higher debris disc fractions in higher UV environments, as the gas is more quickly removed.

\subsection{The extraction of dust mass}
\label{sec:dustExtracted}
An interesting quantity is the total mass in dust extracted from the disc, or similarly the fraction of the initial dust mass that is depleted over the calculation. We know that the total gas mass extracted in each calculation is essentially the initial mass, since all of it is eventually either accreted or stripped in the photoevaporative wind. We however assume in our calculations that there is insignificant loss of solid mass by accretion on to the star once the streaming instability is triggered at the snow-line; moreover at this stage the
dust has grown to a sufficiently large size that it cannot be lost via the wind. Therefore at the end of the calculation only some subset of the initial dust population will have been removed.

Figure \ref{lostDustFraction} shows the percentage of the initial dust mass that is lost over our entire grid of models. This is vastly more sensitive to the initial extent of the disc than the initial disc mass (i.e. there is stronger variation vertically than horizontally in this plot). The fraction of dust lost is almost entirely insensitive to the initial mass. Only once the discs get very high or low mass does it have an appreciable affect on the fraction of dust lost.

In the high mass regime the reason for a smaller fraction of the dust being depleted is simply that photoevaporation is less effective, taking longer to truncate the disc and gain access to more of the dust reservoir. During this time grain growth occurs, and so by the time the disc is truncated most of the dust will be of a size that is not entrained in the wind. In the low mass regime grains start drifting more quickly (equation \ref{equn:adequn}) and so are not entrained in the flow.

\begin{figure*}
	\centering
	\includegraphics[width=18cm]{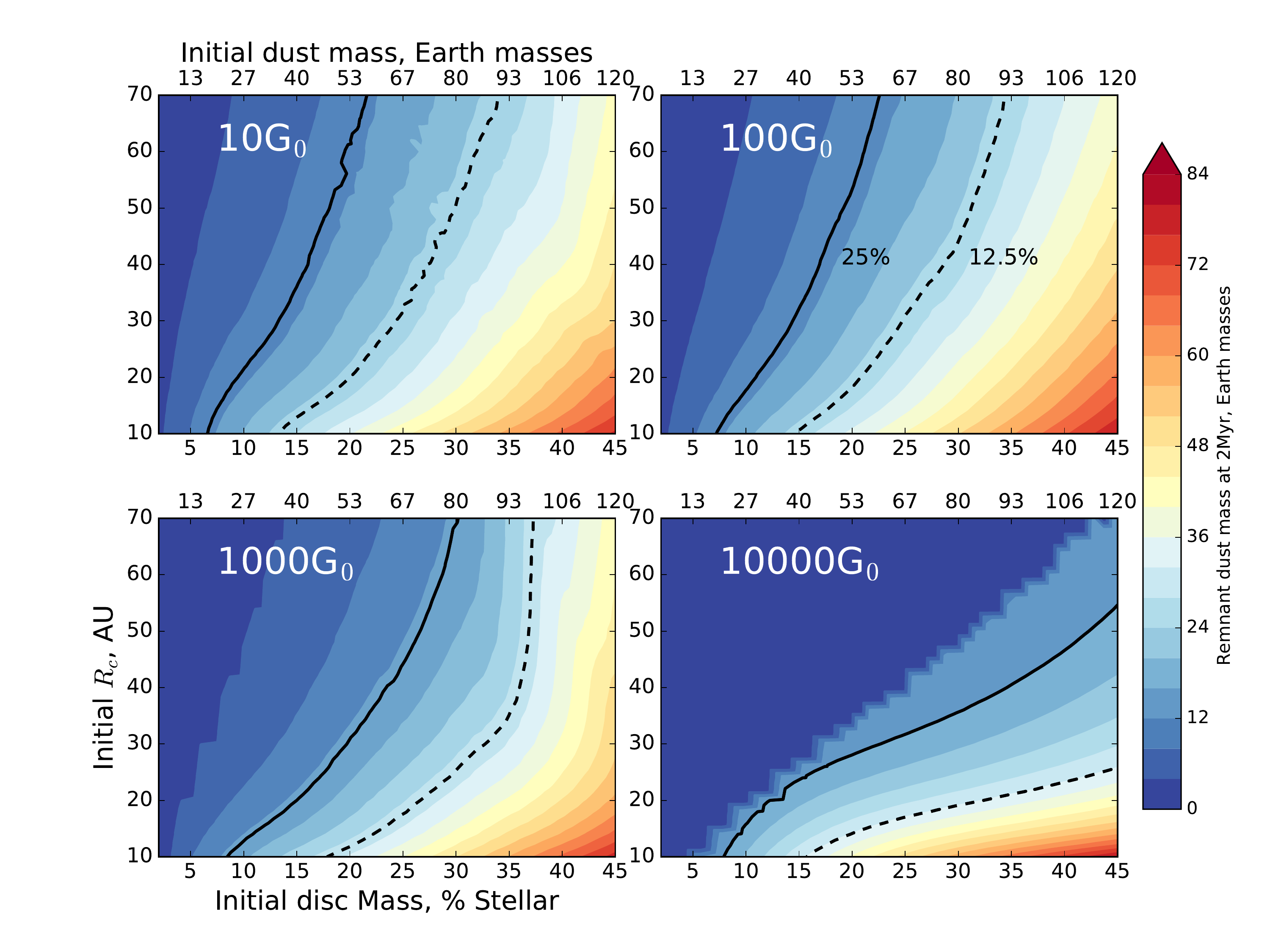}
	\caption{The mass in dust remaining in our grid of discs after 2\,Myr. By this time, dust has stopped being significantly depleted in the photoevaporative wind so this mass is a good representation of what will be available for planet formation (typically, the models reached these plateau masses at much earlier times than this). The solid and dashed contours denote the bondary between models where there is and is not enough mass to produce the Trappist-1 planet with certain efficiencies. That is, to the right of the solid line, conversion of dust into planets with 25\,per cent efficiency would be sufficient, and similarly for the dashed line for an efficiency of 12.5\,per cent. }
	\label{dustmass2Myr}
\end{figure*}

\subsection{Where can a Trappist-1 planetary system be produced?}
\label{sec:where}
For the \cite{2017A&A...604A...1O} planet formation mechanism to operate we require at least 16$M_{\oplus}$ of dust mass at the water snow line in order to facilitate the production of 4$M_{\oplus}$ worth of planets with their assumed 25\,per cent pebble accretion efficiency. In practice more than this will be required because not all of the dust that escapes entrainment in the wind will make it to the water snow line, however this fraction is unknown in our models. 

In Figure \ref{dustmass2Myr} we show the dust mass remaining after 2\,Myr in our grid. Recall that at this time all discs are at the plateau dust mass (and have been for some time) so this amount of dust represents the maximum available reservoir. On these plots we include contours that represent the critical mass for forming the 4$M_{\oplus}$ worth of planets in Trappist-1 assuming different total efficiencies (i.e. the fraction of dust required to make it to the water snow line and end up in planets). The solid contour denotes an efficiency of 25\,per cent (a quarter of the dust remaining after photoevaporation/accretion ends up in planets) and the dashed line an efficiency of 12.5\,per cent.

There are a number of interesting inferences to be drawn from Figure \ref{dustmass2Myr}. Firstly in low-intermediate radiation field environments ($10-1000$\,G$_0$) the dust reservoir available for planet formation is relatively insensitive to the UV field strength (though recall that the gas is efficiently depleted). Above 1000\,G$_0$ this changes, and the initial disc parameters have to be more massive and more compact in order to sustain sufficient dust mass to produce the planets. We remind the reader that although depletion of the dust reservoir doesn't change much between our lower radiation field strengths,  they still deplete at least a factor of two more of the total fraction of dust over 10\,Myr than accretion alone (no photoevaporation) even when accretion is permitted to continue over the full 10\,Myr (see section \ref{sec:focussedEvo}, Figure \ref{fig:totalMassLost}).

We note that this relative insensitivity of the dust mass to the UV field in low-intermediate UV environments is not in conflict with the radial dependence of the dust mass in discs about Sigma Orionis by \cite{2017AJ....153..240A}. They quote a radial variation of the UV field of 8000(d/pc)$^{-2}$\,$G_0$ which gives a UV field of 1280\,$G_0$ at 2.5\,pc, which is conservatively the largest distance that they consider from Sigma Ori. The majority of their discs are well within this distance and hence are exposed to significantly higher UV field strengths. For such UV fields the dust in the discs observed by \cite{2017AJ....153..240A} are expected to be showing signs of radiation field dependent dust mass according to our models (though note that this is complicated since in this paper we are only considering discs about low mass stars). 

Returning to planet formation, the relative insensitivity of the dust mass to the UV up to about 1000\,$G_0$ suggests that Trappist-1 planet formation would be quite resilient to radiation environment in this regime. However in more massive star forming regions such as Orion, there will be regions close to O stars where Trappist-1 planet formation is expected to be more difficult. 

Another major conclusion drawn from these models is that even in low UV environments \textit{the initial mass of the Trappist-1 disc has to be high}. In addition they should also preferably be compact ($R_c < 30$\,AU). For the extremely efficient conversion of 25\,per cent of the remaining solids into planets we require typical initial disc masses around 15-25\,per cent of the stellar. For more conservative efficiencies where 12.5\,per cent of the dust is converted into planets the initial disc mass readily needs to be more like 30-40\,per cent of the stellar. We reiterate that we checked the Toomre Q parameters for these models, and even the most massive are initially stable at all radii. Note that this stability would not hold at similarly high disc to star
mass ratios in the case of more massive stars since gravitational
instability is promoted, for given disc to star mass ratio, by the relatively
low disc aspect ratios in higher mass stars.

\subsection{What fraction of low mass M dwarfs could host a Trappist-1 planetary system}
A key question is to ask what the expected frequency of Trappist-1 like planetary systems would be from our models? Given that our models constrain the types of discs in different environments that can form such a system, if we knew the initial conditions for a large population of discs we could compute what fraction of such discs could produce a Trappist-1 like system -- placing an upper limit on the frequency. Unfortunately the disc initial mass/radius function is highly uncertain. 

The probability distribution of UV fields to which a disc is exposed to is better constrained. The UV field associated with embedded clusters was studied by  \cite{2008ApJ...675.1361F}, who used the initial mass function and cluster size relation to compute the probability that a cluster member is exposed to a given UV field. This assumed that the fraction of stars born in clusters of size $N$ is evenly distributed logarithmically, which is observed up to N$\sim2000$ \citep{2003ARA&A..41...57L}, but assumed to be the case for larger clusters too. \cite{2008ApJ...675.1361F} also considered different extinction levels. Typically the distribution of $P(G_0)$ is approximately a Gaussian, and we used the ``standard'' such distribution from \cite{2008ApJ...675.1361F} which has a median $\log_{10}({G_0})=3.06$ and width 1.13.  For further information on the statistical properties of stellar birth environments, see \cite{2006ApJ...641..504A} and  \cite{2010ARA&A..48...47A}. 

\begin{figure}
	\includegraphics[width=9cm]{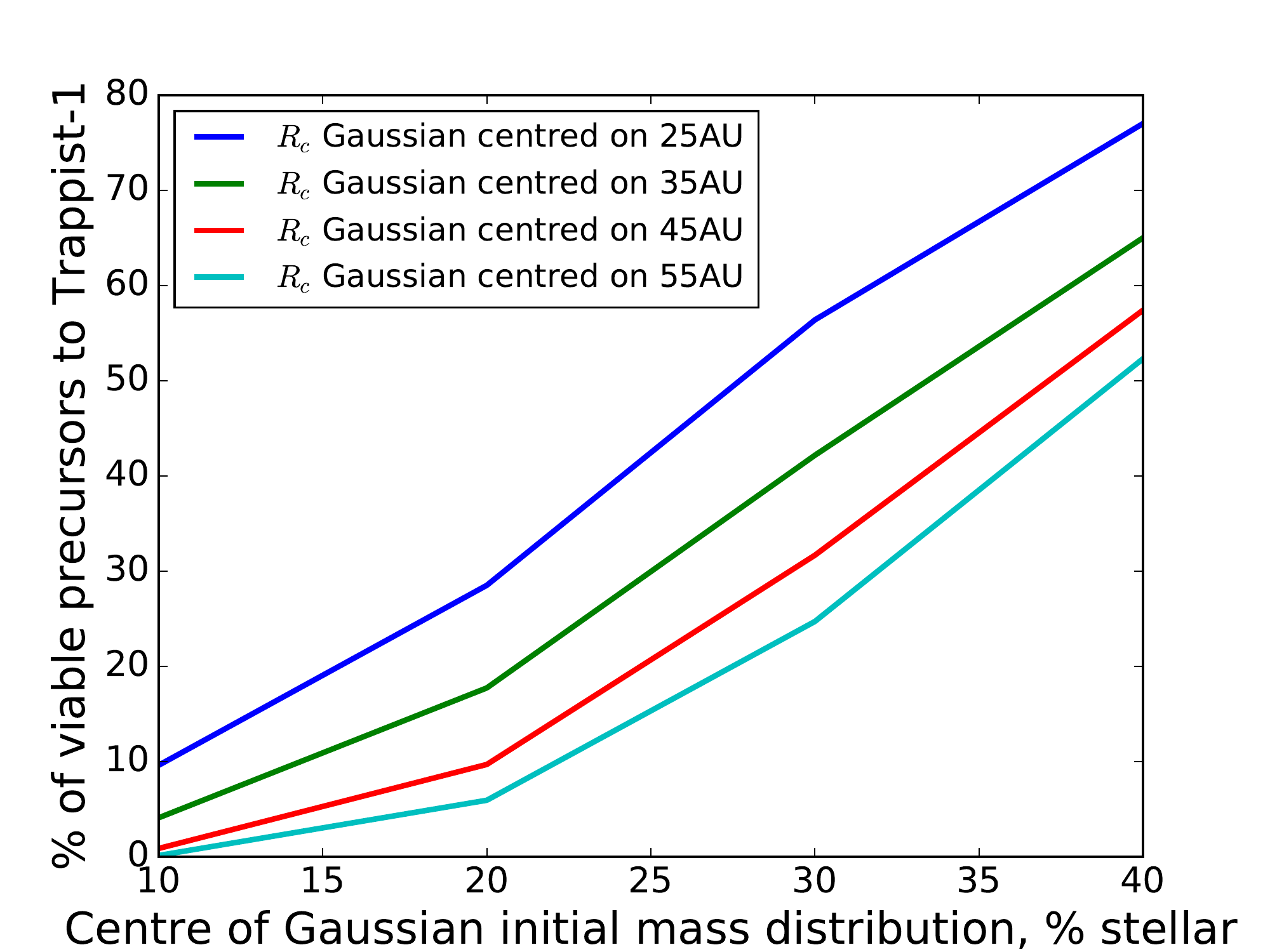}
	\caption{The percentage of Trappist-1 planets able to host a Trappist-1 system for different initial disc parameter distributions. }
	\label{fig:fractionViable}
\end{figure}

Since the initial disc parameters are uncertain, we assume a gaussian distribution of disc masses centred about either 10, 20, 30 or 40\,per cent of the stellar, each of which has a standard deviation of $10$\,per cent. Similarly, the $R_c$ is randomly sampled from a Gaussian centred on either 25, 35, 45, or 55\,AU with a standard deviation of 15\,AU.  This gives a total of 16 initial disc parameter distributions that we randomly sample from to gauge what fraction of M dwarfs from that population could produce Trappist-1 planets. Obviously these populations are artificial (and unconstrained), but they permit us to illustrate what each would yield in terms of Trappist-1 analogue planet populations. 

Figure \ref{fig:fractionViable} shows the percentage of M dwarfs capable of producing a Trappist-1 like planetary system as a function of the peak of the Gaussian of initial disc to star mass ratios, for different disc size distributions. For this plot we assume that 12.5\,per cent of the remaining dust is converted into planets (i.e. there has to be at least 32\,M$_{\oplus}$ worth of dust in the disc).  This fraction is a strong function of the initial disc mass (typically more so than $R_c$, c.f. Figure \ref{dustmass2Myr}). For a mass distribution centred on 10\,per cent of the stellar less then 10\,per cent of M dwarfs are capable of hosting a Trappist-1 planetary system. Much higher initial disc masses, centred on at least 30\,per cent of the stellar, are required for the majority of M dwarfs to host Trappist-1 planets. Better statistics on the fraction of M dwarfs with similar planetary systems might permit us to gain insight into the, highly uncertain, initial conditions of such discs.

\subsection{Could Trappist-1 host an additional higher mass planet?}
\cite{2017arXiv170802200B} recently placed constraints on the upper mass limit of a long period planet in Trappist-1. They found an upper limit of 4.6$M_{\textrm{J}}$ (Jupiter masses) and 1.6$M_{\textrm{J}}$ for a planet with an orbital period of 1 and 5 years respectively. Such a planet would require access to the same limited mass reservoir that we are tracking in our models and hence make it more difficult to produce the confirmed planets. We re-computed the fraction of viable Trappist-1 precursors (as above) requiring an additional tenth of a Jupiter mass of solids go into producing the core of some higher mass planet, under the assumption that the rest of the planetary mass comes from the gas phase. We find that even for our most massive and most compact disc distributions only a very small fraction of the disc population are able to produce both the Trappist-1 planets as well as a the $0.1\,M_{\textrm{J}}$ core for a higher mass planet (irrespective of whether or not there is enough gas for the core to accrete a substantial gaseous atmosphere). We therefore conclude that an additional higher mass, longer period, planet is highly unlikely. If one is detected about Trappist-1 this would imply a very high initial disc mass, it is likely that the longer period planet would be of sub-Jupiter mass (probably only of order another Earth mass)  and such a system would still be expected to be rare for such low mass M dwarfs. 

\subsection{Sensitivity to viscosity}
\begin{figure}
	\includegraphics[width=9cm]{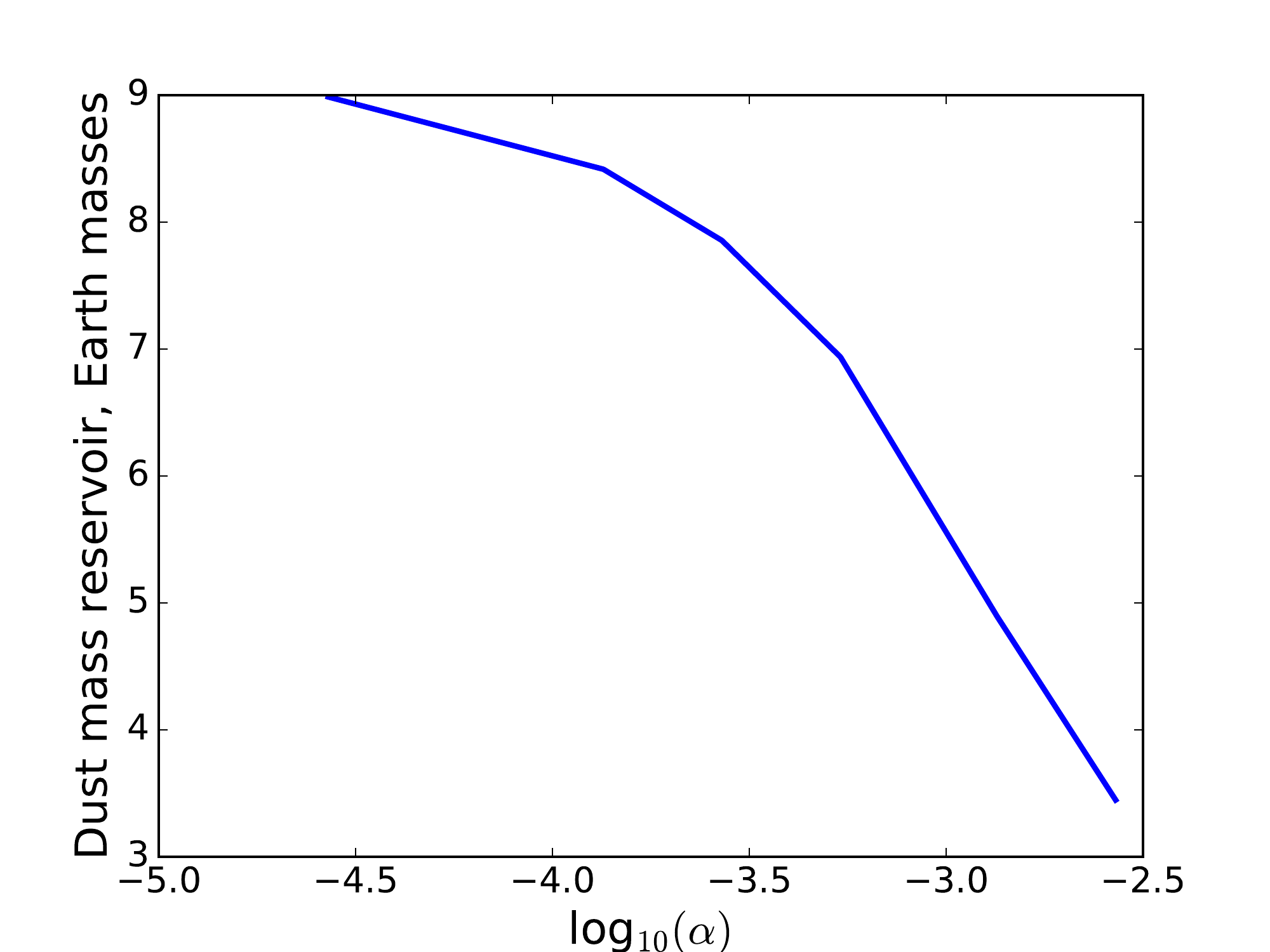}
	\caption{{An illustration of the sensitivity of the remaining dust reservoir to the viscous $\alpha$ parameter, which is set in our models by a combination of the disc parameters and initial accretion rate. This is for a single disc setup with initial disc mass 20\,per cent the stellar and $R_c=50$\,AU.  The lowest $\alpha$ value is that resulting from our main grid for this model.} }
	\label{fig:alphaScale}
\end{figure}

{The viscosity in our models is set by the disc parameters and accretion rate. It hence varies across the disc parameter space since we assume the same initial (observationally motivated) accretion rate. Recall from equation \ref{equn:alpha} that for a disc mass 1\,per cent of the stellar and $R_c=50\,$AU $\alpha\sim10^{-3}$, with $\alpha$ decreasing for higher mass, more compact discs. The $\alpha$ viscosity parameter is therefore quite low across much of our parameter space. We therefore ran additional calculations where we increased the accretion rate (and hence $\alpha)$. Figure \ref{fig:alphaScale} shows the dust reservoir for planet formation as a function of $\alpha$ for a disc with initial mass 20\,per cent the stellar, $R_c=50$\,AU that is irradiated by 100\,G$_0$. In line with \cite{2013ApJ...774....9A} we find that a higher $\alpha$ leads to more effective evaporation of the disc, including of the dust mass reservoir. Overall then, unless viscosity is extraordinarily low, the high disc masses required by our models to form the Trappist-1 planets cannot be circumvented by changes to the viscosity. }

\section{Summary and conclusions}
We have used a large grid of viscous evolutionary models, including accretion onto the star and external photoevaporation, to study the evolution of discs that might be precursors to the famous Trappist-1 system. We aim to probe the dust and gas mass content of such discs over time in different radiation environments, and any consequences that this has for planet formation. We draw the following main conclusions from this work. \\

\noindent 1. Photoevaporation of the gas from  a Trappist-1 precursor disc is effective, owing to the low stellar mass and hence weak boundedness of the material at the disc outer edge. Intuitively, more massive, compact, weakly irradiated discs retain their gas reservoir for longer than low mass, extended, strongly irradiated discs.  \\

\noindent 2. Once grains reach a critical size \citep[as shown by][]{2016MNRAS.457.3593F} they are no longer entrained in the photoevaporative flow. Evolving the maximum grain size following \cite{2012A&A...539A.148B} results in a front that propagates quickly outward through the disc, interior to which grains are sufficiently large that dust is resilient to the effects of photoevaporation. Photoevaporation hence only has a short period of time in which to deplete the disc of dust (see also appendix \ref{sec:appendix1}). We demonstrate that radial drift dominates fragmentation in setting the maximum grain size across our parameter space.   \\

\noindent 3. Although dust quickly becomes resilient to external photoevporation, gas continues to be depleted. The solids-to-gas ratio hence increases over time (at a rate dependent on the disc properties and incident UV field strength, see Figure \ref{fig:detailedDustToGas}). Eventually the mass in solids dominates over the gas, and shortly after this the gas disc is cleared completely. The gas disc lifetime is hence synonymous with the time at which the disc would be identified as a debris disc. For a distribution of initial disc parameters our models expect a mixture of debris and protoplanetary discs in a given young stellar cluster, even for a single UV field environment (at least for low mass stars). \\

\noindent 4. Although dust rapidly becomes resilient to photoevaporation, a significant fraction of the initial dust reservoir can still be removed early on. We find that high initial disc masses are required in order to form the observed 4\,M$_{\oplus}$ worth of Trappist-1 planets. For example, if the dust reservoir resilient to photoevaporation is converted into planets with 25\,per cent efficiency initial disc masses of 15--25\,per cent the stellar are required in 10-$10^3$\,G$_0$ environments. For 12.5\,per cent efficiency the initial disc mass requrired is pushed towards 30--40\,per cent the stellar. In 10$^4$\,G$_0$ environments only the most compact (R$_c<30$\,AU) discs are capable of producing the Trappist-1 planets. Note that we confirmed that such high fractional disc masses are Toomre $Q$ stable. \\

\noindent 5. Even very weak UV fields strip substantially more dust (by about a factor of 2) than 10\,Myr of depletion through accretion only. \\

\noindent 6. We estimate the fraction of Trappist-1 mass M dwarfs capable of hosting a similar planetary system by randomly sampling Gaussian distributions of initial disc parameters.  As above, distributions centered on high masses ($>30$\,per cent the stellar) are required for $\sim50$\,per cent of such M dwarfs to be able to host similar planetary systems, at least for an efficiency with which solids are converted to planets of 12.5\,per cent. \\

\noindent 7. It is almost impossible for an additional higher mass, longer orbital period planet to be produced around Trappist-1. None of our models can accommodate an additional 0.1 Jupiter mass planetary core unless the efficiency with which grains are converted into planets is unrealistically high, and also only for our most compact and massive discs. We hence conclude that an additional longer period high mass planet is very unlikely, and if one were to exist would probably be sub-Jupiter mass.   \\

\section*{Acknowledgements}
{We thank the anonymous referee for their insightful comments.}
We also thank Fred Adams, James Owen and Hugh Jones for useful discussions. 
TJH is funded by an Imperial College London Junior Research Fellowship. CJC acknowledges support from the
DISCSIM project, grant agreement 34113, 7 funded by the European Research Council
under ERC-2013-ADG.
This work was partly developed during and benefited from the MIAPP ``Protoplanetary discs and planet formation and evolution'' programme. The photoevaporation models in this paper were run on the COSMOS
Shared Memory system at DAMTP, University of Cambridge operated
on behalf of the STFC DiRAC HPC Facility. This equipment is
funded by BIS National E-infrastructure capital grant ST/J005673/1
and STFC grants ST/H008586/1, ST/K00333X/1. DiRAC is part of
the National E-Infrastructure.

\bibliographystyle{mn2e}
\bibliography{molecular}

\appendix

 \section{Analytic estimate of the grain entrainment fraction}
\label{sec:appendix1}
For grain growth and entrainment in the wind as discussed in section \ref{sec:dustStrip} we can solve for the dust mass loss rate over time as a function of disc outer radius in the disc as grain growth proceeds (i.e. by combining and integrating equations \ref{equn:grainEntrainMax}-\ref{equn:amaxOverTime})
 \begin{equation}
 \dot{M}_{dw} = \delta \dot{M}_w^{3/2}\left(\frac{v_{th}}{4\pi \mathcal{F} G M_* {\rho_g} a_{min}}\right)^{1/2}\exp\left[\frac{-\delta \left(GM_*\right)^{1/2}t}{2R^{3/2}}\right].
 \end{equation}
 
  An interesting quantity is the ratio of the dust mass loss rate to that one would expect for perfect entrainment, i.e. $\Xi=\dot{M}_{dw}/(\delta\dot{M}_w)$, where $\delta$ is the initial dust-to-gas mass ratio. Assuming that the scale height is $c_s/\Omega$ this is
\begin{multline}
	\Xi = \left(\dot{M}_w\,M_\odot\,yr^{-1}\right)^{1/2} \left[\left(\frac{8^{1/2}}{4\pi^{3/2}G^{1/2}M_*^{1/2}{\rho_g}a_{min}}\right)\right. \times \\
	\left. R^{-3/2}\left(\frac{k_B T R^3}{\mu m_H G M_*} + R^2\right)^{1/2}\right]^{1/2}\exp\left(-\frac{\delta G^{1/2}M_*^{1/2}t}{2R^{3/2}}\right).
	\label{entr_fraction}
\end{multline}
From our grid (Figure \ref{MdotGrid}) we know that the gas mass loss rate is more sensitive to radius than surface density. We can therefore fit some representative $\dot{M}_w(R)$ and feed it into equation \ref{entr_fraction} to determine how the dust fraction entrained evolves as a function of time and radius. To facilitate this we fitted a fifth order polynomial to the mass loss profile of a disc with initial mass 20\,per cent the stellar
\begin{equation}
	\log_{10}\left(\dot{M}_w\right) = p_1 + p_2x +  p_3x^2 +  p_4x^3 +  p_5x^4 + p_6x^5
\end{equation}
the paramters for which in the case of a 10 and 1000\,G$_0$ field are given in Table \ref{tab:parameterisedMdot}.

\begin{table}
	\caption{Parameters for a polynomial fit to the photoevaporative mass loss rate as a function of disc outer radius from a Trappist-1 precursor disc in 10 and 1000\,G$_0$ environments. These are specifically tailored for a disc with initial mass 20\,per cent of the stellar, though note that compared to the disc outer radius the mass loss rate is \textit{relatively} insensitive to the surface density and hence mass. }
	\label{tab:parameterisedMdot}
	\centering
	\begin{tabular}{c l l l l }
	\hline
	& $10G_0$ &  $1000G_0$& \\
	\hline
	$p_1$ & -13.8842  & -12.4687\\
	$p_2$ &  0.363892 & 0.424697 \\
	$p_3$ & -0.00557322 & -0.009659\\
	$p_4$ & 3.3816e-051 & 0.00010268\\
	$p_5$ &-6.21243e-08   & -5.25644e-07 \\
	$p_6$ & -5.99367e-11 & 1.04366e-09\\	
	\hline
	\end{tabular}
\end{table}
 
\begin{figure}
	\vspace{-0.5cm}

	\hspace{-0.5cm}
	\includegraphics[width=9cm]{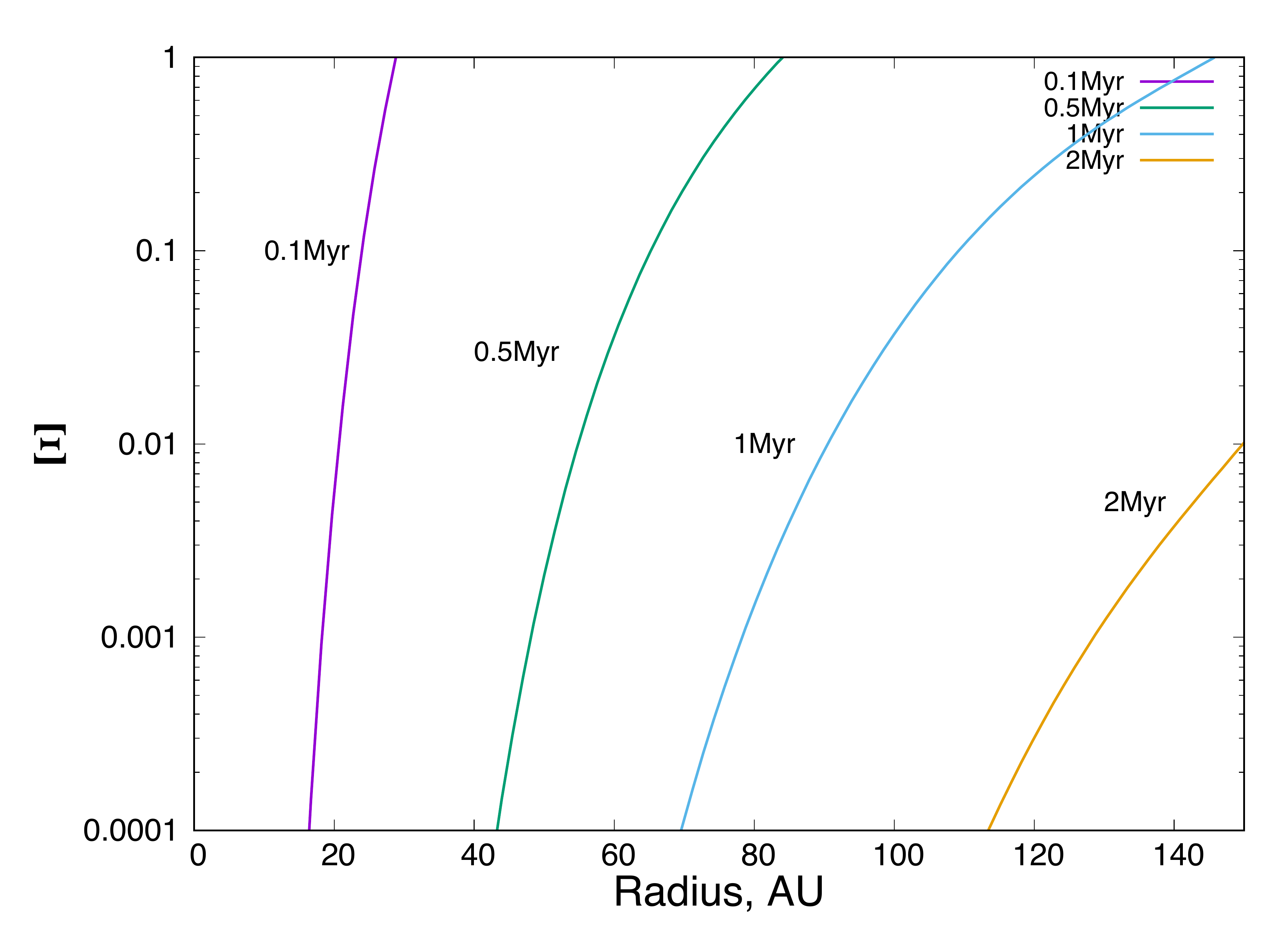}

	\vspace{-0.3cm}
	\caption{The fraction of dust entrained in a photoevaporative wind as a function of disc outer radius, with each line denoting different times. That is, at 0.1\,Myr dust within about 20\,AU has grown to sufficient size that it won't be entrained in the wind. By 2\,Myr photoevaporative stripping of dust would be inefficient at all radii.  }
	\label{fig:analyticGrainFracEntrained}
\end{figure}

Combining the above, we can estimate the fraction of the dust entrained in the wind as a function of radius at different times, an example of which is shown in Figure \ref{fig:analyticGrainFracEntrained} for the 1000\,G$_0$ radiation field parameterisation from Table \ref{tab:parameterisedMdot}. This shows that after 0.1\,Myr essentially all dust within about 20\,AU has grown to sufficient size that it is impervious to photoevaporative stripping. By 0.5\,Myr all dust within $\sim65$\,AU is resilient to photoevaporation and by 2\,Myr essentially the whole disc retains its dust.

These analytic estimates are powerful tools for quickly estimating the dust stripping and mass reservoir for planet formation for a given low mass M-dwarf disc, without requiring additional numerical models.  This kind of analysis could be extended to other stellar systems with future more comprehensive/dedicated mass loss rate grids.

\section{Notation summary}
\label{sec:notationSummary}

\begin{table}
	\caption{A summary of the key notation used in this paper. }
	\label{tab:symbols}
	\centering
	\begin{tabular}{c l}
	\hline
	\hline	
	Symbol & Description  \\
	\hline
	$\dot{M}_a$ & Gas mass accretion rate \\
	$\dot{M}_w$ & Gas mass loss rate in photoevaporative wind \\
	$\dot{M}_{dw}$ & Dust mass loss rate in photoevaporative wind \\
	\hline 
	$M_*$& Stellar mass\\	
	$M_g$& Disc gas mass\\
	$M_d$& Disc dust mass\\
	$m_p$ & Planetary mass \\
	\hline
	$a_{\textrm{entr}}$ & Maximum grain size entrained in photoevaporative wind \\
	$a_{\textrm{max}}$ & Maximum grain size at a given radius \\
	$a_{\textrm{min}}$ & Minimum grain size at a given radius \\	
	$a_d$ & Size of grains at which radial drift occurs\\
	$a_f$ & Size of grains at which fragmentation occurs\\
	$\rho_g$ & Grain mass density \\
	$\delta$ & Dust-to-gas mass ratio \\
	$v_f$ & Fragmentation velocity\\
	\hline
	$\Sigma$ & Surface density\\
	$\Omega$ & Angular frequency \\
	$T$ & Temperature \\
	$c_s$ & Sound speed\\
	$v_{\textrm{th}}$ & Mean thermal speed of gas particles\\	
	$H_d$ & Scale height at disc outer edge\\	
	$R_d$ & Disc outer radius \\
	$R_c$ & Scaling radius (equation \ref{SigmaProf})\\
	$\Sigma_c$ & Surface density at scaling radius\\
	$\mathcal{F}$& Solid angle subtended by disc at outer edge \\
	$Q$& Toomre stability parameter\\
	$\alpha$ & viscosity parameter \\
	$\nu$ & viscosity \\
	\hline
	$\Xi$ & Instantaneous fraction of possible dust entrained in wind \\
	$\eta$ & Planet formation efficiency (planets-to-solids mass ratio)\\
	\hline
	\hline
	\end{tabular}
\end{table}

\label{lastpage}

\end{document}